\documentclass[a4paper,11pt]{article}
\pdfoutput=1 
\usepackage{jinstpub}
\usepackage{fancyhdr}
\usepackage[T1]{fontenc} 
\usepackage{subcaption}
\usepackage{xcolor, soul}
\usepackage{amsmath}
\numberwithin{equation}{section}
\sethlcolor{red}
\title{\boldmath Neural networks for boosted di-$\tau$ identification}

\author[a]{Nadav Tamir,}
\author[a]{Ilan Bessudo,}
\author[a]{Boping Chen,}
\author[a]{Hely Raiko,}
\author[a]{and Liron Barak}

\affiliation[a]{Tel Aviv University,\\Tel Aviv, Israel}

\emailAdd{nadav.michael.tamir@cern.ch}
\emailAdd{ilanibm@hotmail.com}
\emailAdd{boping.chen@cern.ch}
\emailAdd{helyraiko@mail.tau.ac.il}
\emailAdd{lironbarak83@gmail.com
}

\abstract{We train several neural networks and boosted decision trees to discriminate fully-hadronic boosted di-$\tau$ topologies against background QCD jets, using calorimeter and tracking information. Boosted di-$\tau$ topologies consisting of a pair of highly collimated $\tau$-leptons, arise from the decay of a highly energetic Standard Model Higgs or Z boson or from particles beyond the Standard Model. We compare the tagging performance for different neural-network models and a boosted decision tree, the latter serving as a simple benchmark machine learning model.}

\begin{document} 
\maketitle
\newcommand\blfootnote[1]{
    \begingroup
    \renewcommand\thefootnote{}\footnote{#1}
    \addtocounter{footnote}{-1}
    \endgroup
}
\blfootnote{The code used to obtain the results presented in this paper is available on \href{https://github.com/ntamir/BoostDiTau_ML}{GitHub}.}
\flushbottom
\newpage
\section{Introduction}

In current high-energy physics (HEP) experiments such as ATLAS and CMS, final-state particles can often be produced with significant Lorentz boosts, resulting in signatures localized in a small area inside the detector. These boosted final-state topologies provide a unique window to study phase-space regions less explored by traditional analyses -- both in Standard Model (SM) measurements and in searches for physics Beyond the Standard Model (BSM).

Since the angular separation between decay products of a resonance particle -- henceforth denoted as "X" --  is proportional to the ratio between the resonance's mass and its momentum, the onset of the boosted regime will occur at lower transverse momenta ($p_T$) for lighter resonances. For example, the Standard Model (SM) Higgs boson decaying into a particle-antiparticle pair of $b$-quarks \cite{PERF-2017-04,CMS-HIG-17-010} or $\tau$-leptons \cite{HDBS-2019-22} will have a boosted phase space at large Higgs $p_T$, a feature used in searches for production of heavy BSM resonances \cite{HDBS-2018-17,CMS-B2G-20-009}. Alternatively, the boosted decay regime of a light ($m_X\lesssim\frac{m_H}{2}$) resonance decaying to boosted SM particles, occurring at lower characteristic $p_T$s, can be the target \cite{HDBS-2018-47}.

Among the many possible boosted topologies consisting of various SM particles, an interesting channel is the $\tau^+\tau^-$ final state. This final state could be sensitive to BSM scenarios, for instance, the "hidden sector", involving couplings to SM particles that are proportional to their mass~\cite{Curtin:2013fra,Robens:2015gla,Robens:2016xkb,Robens:2019kga,Bauer:2017ris,Shrock:1982kd,Strassler:2006im,Schabinger:2005ei,Patt:2006fw}.

Reconstructing and tagging boosted particles in a hadronic topology ("boosted-Jet tagging") is a rather challenging task, limited by experimental sensitivity. The performance of "standard" reconstruction and tagging techniques tends to degrade with decreasing angular separation and dedicated methods are required, most commonly making use of the internal structure of a "Large-Radius-Jet" (LRJ) to tag the constituent boosted particles~\cite{JETM-2018-03,CMS-JME-18-002}. In particular, an algorithm for reconstructing and tagging single hadronically decaying $\tau$-leptons ($\tau_h$) may fail to reconstruct or identify a $\tau_h$-pair when the decay products overlap, degrading the identification efficiency.
In this case, the highly collimated $\tau_h^+\tau_h^-$ pair, referred to as the boosted di-$\tau$ object, requires a dedicated tagging algorithm that distinguishes it from the most common background noise in proton-proton collisions -- the copious numbers of hadronic jets arising from QCD interactions~\cite{HDBS-2019-22}.

Improvements in particle tagging -- of which jet tagging is just one case -- are often achieved through application of techniques from the realm of machine learning (ML), using methods of Deep Learning (DL) involving artificial neural-networks (NN's)~\cite{jain1996artificial,hassoun1995fundamentals}. The ability of DL-based methods to deal with very high-dimensional data and their flexibility in handling different structures and correlations in data allow them to learn useful representations of the data as they are trained, giving rise to improved performance over more traditional Multi-Variate-Analysis (MVA) and cut-based methods. Better tagging algorithms are critical in the success of current and future HEP experiments' goals -- both for improving sensitivities in offline analysis~\cite{Guest:2018yhq,Albertsson:2018maf,Radovic:2018dip,Bourilkov:2019yoi,Schwartz:2021ftp,Karagiorgi:2021ngt,DiTau_ML_mass,DL_for_HTT_mass} and for designing more efficient triggers for online operation~\cite{Neu:2023sfh,Bhattacherjee:2023evs,ATLAS:2023gog,Clerbaux:2020ttg,Migliorini:2021fuj,Yaary:2023dvw}. Throughout run-2 of the LHC, for example, both ATLAS and CMS experiments implemented NN-based tagging algorithms as their new standard for jet-tagging tasks~\cite{JETM-2018-03,CMS-JME-18-002}.
 
In this paper, we present a study comparing the performance (in terms of background rejection) of four ML models designed for tagging boosted di-$\tau$ objects originating in the decay of a light resonance. Two of the models use hand-engineered "high-level-features" calculated using combinations of jet- and track-level properties, while the other two use only the raw "low-level-features" taken directly from the underlying constituents. The studied ML models hence differ in their use of available object features and representations of the underlying data, and as such cover different approaches to tackling the classification problem.
\section{MC sample simulation and selection}
This section summarizes the sample generation and the high- and low-level feature calculation used subsequently for training and evaluating the various ML models. It first introduces the detail of the simulated sample, followed by the definition of the di-$\tau$ object, the description of the low-level variables and the definitions of the higher-level variables based on the di-$\tau$ object.
\subsection{Simulated sample}
Simulated samples of a 2HDM pseudoscalar boson production (denoted as X) in association with a top-antitop quark pair (henceforth abbreviated as "$t\bar{t}X$") from proton-proton collisions are used as the signal sample for the studies. The X is set to decay into two $\tau$'s, which are subsequently set to decay hadronically. The X mass is set to $m_{X}=20$ GeV. The SM $pp\rightarrow t\bar{t}$ production process is used as the background sample. The $t\Bar{t}$ pairs in both signal and background sample are set to either a semi-leptonic or di-leptonic decay (with only electrons and muons considered).
The $t\bar{t}X$ and $t\bar{t}$ samples are simulated at a centre-of-mass energy of 13 TeV, using MadGraph5~\cite{Alwall:2014hca} (v2.6.7). Parton shower, hadronization, and underlying event effects are accounted for with Pythia8~\cite{Sjostrand:2014zea}. Detector simulation and reconstruction are done with Delphes~\cite{Ovyn:2009tx,deFavereau:2013fsa} v3.5.0. A customized CMS datacard is used for the simulation~\cite{Pol_2022,delphesconfig}.

\subsection{Di-$\tau$ reconstruction and selection}
\label{sec:ditaureconstruction}
\renewcommand{\thefootnote}{\fnsymbol{footnote}}
The reconstruction for our di-$\tau$ object starts from reconstruction of two types of jets -- the "seed" LRJ's whose area is assumed to contain both $\tau_h$'s signatures, and "sub-jets" (SJ) whose area is assumed to contain a single $\tau_h$ signature. Both types are reconstructed from calorimeter energy deposits using the anti-$k_t$ algorithm with a radius parameter $R=1.0$ for the LRJ and $R=0.2$ for the sub-jets~\cite{Cacciari:2008gp,Fastjet}. 
To be considered a di-$\tau$ object, a LRJ is required to have at least two sub-jets within its area, each with at least one track within its respective area. The "constituents" of the di-$\tau$ are thus the LRJ seed and the sub-jets, calorimeter-cluster and track sets associated with the area of the LRJ. The area of the LRJ between the sub-jets (containing energy deposits not associated with any sub-jet) is then considered as the "isolation region", and used in later background rejection. Each sub-jet is also assigned a "core region", a cone of $\Delta R<0.1$ around its axis\footnote[4]{$\Delta R=\sqrt{(\Delta\eta)^2 + (\Delta\phi)^2}$ ~is a measure of angular distance used in HEP experiments}. A "signal" di-$\tau$ object is defined by a geometric matching between each of its two leading (in $p_T$) sub-jets to a generator-level $\tau_{h}$ with $\Delta R(\textrm{\scriptsize{SJ}~,~} \tau_h)<0.2$.

\subsection{Input features}
\label{highlevel}
Two types of datasets are generated from the samples. One contains the low-level features -- the aforementioned di-$\tau$ constituents -- while the other contains high-level features only (each defined by a single value for a given di-$\tau$ object).

For the low-level dataset, each di-$\tau$ is associated with the $\{E_T,\eta,\Phi\}$ of all its constituents. In addition, calorimeter deposits are associated with a value of $-0.5$ if they originate in an ECAL and $0.5$ if they originate in an HCAL deposit, while tracks have their impact parameter ($d_0$) associated as an additional feature.

The high-level features used as inputs are listed below, where the notation "(sub)lead" represents the (sub)leading (in $p_T$) sub-jet inside the large-R jet:
\begin{enumerate}
    \item $\mathbf{n_{isotracks}}$ : The number of tracks associated to the isolation region.
    \item \textbf{Sub-jet energy fraction ($\mathbf{f_{subjet}^{(sub)lead}}$):} The ratio between the transverse momenta of the sub-jet and the seed jet.
    \begin{equation}
    f_\text{subjet}^\text{(sub)lead}=\frac{p_T^\text{(sub)lead}}{p_T^\text{seed}}\label{eq:f}
    \end{equation}
	\item $\mathbf{R_{isotrack}}$: The $p_T$-weighted sum of track distances from the sub-jet axis, for isolation-region tracks found inside a cone of $\Delta R<0.4$ around the sub-jets. This definition means the variable considers tracks within an "isolation annulus" of $0.2<\Delta R<0.4$ around the sub-jets. A value of zero is assigned if no tracks are found.
    \begin{equation}
    R_\text{isotrack} = \frac{\sum_\text{sub(lead)}\sum_{i}^{\Delta R_i<0.4}p_{T,i}^\text{isotrk}\Delta R_i}{\sum_\text{sub(lead)}\sum_{i}^{\Delta R_i<0.4}p_{T,i}^\text{isotrk}}\label{eq:Riso}
    \end{equation}
	\item $\mathbf{R_{max}^{(sub)lead}}$: The maximal $\Delta R$ of an associated track to the sub-jet axis.
	\item \textbf{Weighted core track distance ($\mathbf{R_{core}^{(sub)lead}}$):} Defined for a given sub-jet, this is the $p_T$-weighted sum of track distances from the sub-jet axis, for tracks found inside the core cone of the sub-jet. A value of zero is assigned if no tracks are found inside the core cone.  
    \begin{equation}
    R_\text{core}^\text{(sub)lead} = \frac{\sum_{i}^{\Delta R_i<0.1}p_{T,i}^\text{trk}\Delta R_i}{\sum_{i}^{\Delta R_i<0.1}p_{T,i}^\text{trk}}\label{eq:Rcore}
    \end{equation}
	\item $\mathbf{R_{track}}$: $p_T$-weighted sum of track distances from the sub-jet axis, for tracks found inside a cone of $\Delta R<0.2$ around the sub-jets.
    \begin{equation}
    R_\text{track} = \frac{\sum_\text{sub(lead)}\sum_{i}^{\Delta R_i<0.2}p_{T,i}^\text{trk}\Delta R_i}{\sum_\text{sub(lead)}\sum_{i}^{\Delta R_i<0.2}p_{T,i}^\text{trk}}\label{eq:Rtrack}
    \end{equation}
	\item $\mathbf{f_{track}^{(sub)lead}}$: Ratio between the highest $p_T$ track inside a sub-jet, and the respective sub-jet $p_T$.
	\item $\mathbf{Log(m_{tracks}^{(sub)lead})}$: Logarithm of the invariant mass calculated from the 4-momenta of tracks associated with the given sub-jet.
	\item $\mathbf{|d_{0,lead-track}^{(sub)lead}|}$: The absolute value of the impact parameter of the leading track associated with the appropriate sub-jet.
	\item $\mathbf{\Delta R(lead,sub-lead)}$: Angular separation between the two leading sub-jets.
\end{enumerate}
A principal component analysis (PCA) decomposition was performed on this set of input features, to study whether the classification task may benefit from this form of dimensionality reduction and to examine the separation achievable via this method. The PCA decomposition did not lead to any obvious separation between the distributions of signal and background, and no particularly striking clustering patterns emerged as a result of its application. The PCA decomposition results, however, provide no insight into nonlinear relations between the high-level input features. Therefore, a t-SNE~\cite{tSNE_paper} and a UMAP~\cite{umap_paper} decomposition were also performed to explore these types of relations. Compared to the PCA results, both of these manifold learning methods yielded more distinct clustering features, and a single component with stronger discriminating power. Further details on the decomposition studies are given in Appendices~\ref{app:PCA} and ~\ref{app:Manifold}.

\section{ML Models and Dataset Processing}
The goal of the various tagging algorithms tested is to classify the real ("signal") boosted di-$\tau$ objects from the fake background originating in QCD jets. It is treated as a supervised learning problem, thus the signal and background samples are given class labels to indicate their type, which are used as the target for prediction in the NN models. 

\subsection{Classification Models}
We trained and evaluated the performance of four tagging models on the aforementioned signal and background samples, three of which were NN-based models. Detailed information of the hyper-parameters used for each of the algorithm can be found in Appendix~\ref{app:BDTNN}.

For the NN's, dropout layers are used after each linear layer, as they were found to improve both overall performance and generalization power of all models while acting also as a regularization method. Models including skip-connections were also tested, but were not found to be more effective. All our NN implementations use Parameterized Rectified Linear Unit (PReLU) activation functions for the hidden layers, and a sigmoid activation function for the output -- yielding a continuous score on the interval $\left(0,1\right)$. The datasets used for training, validation and testing of the model all have an approximately equal number of signal and background events. Binary Cross-Entropy with L2 regularization is used as the loss function in all cases. The values of batch size, learning rate, L2 regularization weight normalization and dropout fraction hyper-parameters for the models reported in this paper are chosen via an optimization study conducted using Optuna~\cite{optuna_2019}. The choice of network dimensions was derived in part from a manual optimization process, and in part from a similar optimization study via Optuna.

\subsubsection{BDT}
The first model studied is a Boosted Decision Tree (BDT) algorithm~\cite{roe2005boosted}, which has been successfully used in many HEP analyses~\cite{FTAG-2019-07,CMS-BTV-16-002}. This algorithm uses the high-level-feature dataset described in Section~\ref{highlevel}, and iteratively creates a "forest" -- an ensemble of decision trees, with each consecutive tree specializing in classifying events the previous trees failed to correctly predict. Each tree acts as a weak classifier predicting a single class label ($-1$ for background and $1$ for signal), and the weighted average of all trees in the forest thus results in a continuous score on the interval $\left[-1,1\right]$.

\subsubsection{DNN/MLP}
The second model is the deep neural network (DNN) -- a multi-layer perceptron (MLP) with fully connected layers, using the same high-level-feature dataset as the BDT. Unlike the BDT, a DNN has the ability to learn both linear and non-linear relationships between the input variables, via the nonlinearities introduced by the activation function. It is also widely used in many HEP analyses~\cite{FTAG-2019-07,CMS-BTV-16-002}.   

\subsubsection{CNN \label{subsubsec:CNN}}
The third model is a convolutional neural network (CNN), which is widely used for image classification \cite{albawi2017understanding,gu2018recent} and whose successes helped herald the recent decade's advances in the field of computer vision \cite{alexnet}. We use the low-level-feature dataset for the CNN, of which only the tracks and calorimeter deposits are considered and only the $\{E_T,\eta,\Phi\}$ are used as input features.

Usage of a CNN for this case is motivated by the observation that the particle detector's barrel-region coordinate space can be unfolded onto a 2D projection -- similar to an image from a camera -- and by mapping the ECAL, HCAL and tracks $E_{T}$ into three individual copies of this $\eta$-$\phi$ surface, one ends up with an image-like distribution. The $\eta$ and $\phi$ coordinates correspond to X and Y axis locations, and the $E_{T}$ values for the different input types correspond to the pixel values in the RGB filters of the image. This representation enables the use of image identification algorithms in the jet classification task. 

\subsubsection{DeepSet}
The fourth architecture is a DeepSet (DS), which by construction is built to handle cases of variable-sized inputs while respecting permutation invariance among the inputs~\cite{zaheer2017deep}. DS architectures update local set element representations via application of a permutation-invariant aggregation operator on all set elements, which allows global set information to propagate to the individual elements' representations. We use the low-level-feature dataset for the DS, including all constituents and their features. 

\begin{figure}[h!] 
    \centering
    \includegraphics[width=1.0\linewidth]{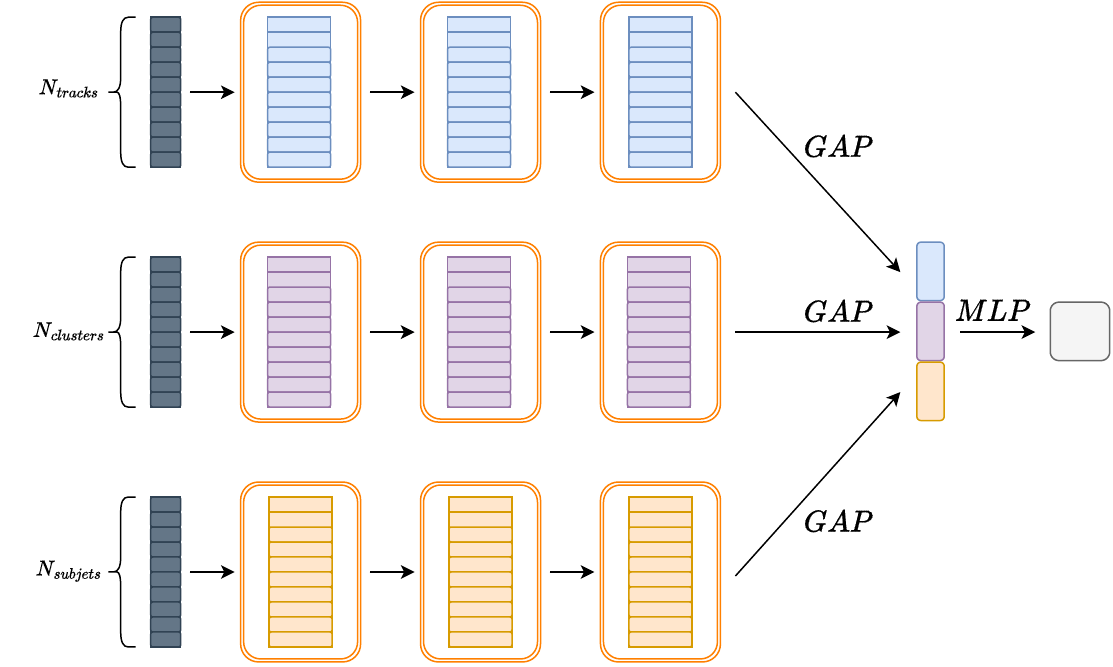}
    \caption{Illustration of the DeepSet implementation. Input sets pass through DS blocks (orange frames), summed over the set elements via Global Attention Pooling (GAP) layers, then passed through a classification MLP to obtain the NN score.}
    \label{fig:DSNN}
\end{figure}

\begin{figure}[h!] 
    \centering
    \includegraphics[width=1.0\linewidth]{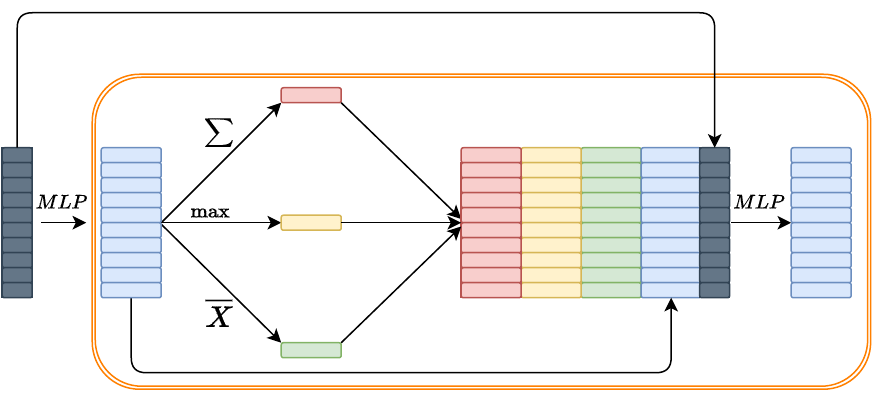}
    \caption{Illustration of a single DeepSet block. Three aggregated global representations are formed from the latent element representations, replicated and concatenated with the latent representations and input features, then updated via a MLP.}
    \label{fig:DSblock}
\end{figure}

In the DS implementation, each input type is treated separately. First, an initialization MLP creates a latent representation for the set elements. It is then updated through three DS blocks. In each block the mean, sum and maximum of the latent representation across the set elements is calculated. These global, permutation-invariant representations are subsequently concatenated with the latent representation and the input features, then passed through a shared-weight MLP to obtain an updated latent representation. After the three DS blocks, an attention-pooling layer is applied, yielding a single global latent representation for each of the input sets. The three resultant global representations are concatenated and passed through a MLP to obtain the final network output (the classification score). An illustration of the overall DS structure is given in Figure~\ref{fig:DSNN}, with the inner structure of each DS block depicted in Figure~\ref{fig:DSblock}. 
\newpage
\subsection{Dataset Pre-Processing}
\raggedbottom
The four different algorithms presented above each treat the data differently in terms of structure and in terms of what the data actually represents, and as such, different pre-processings must be applied before the data can be delivered to a given algorithm.
\begin{figure}
    \centering
    \begin{subfigure}[ht]{0.35\paperwidth}
        \centering
        \includegraphics[width=0.35\paperwidth]{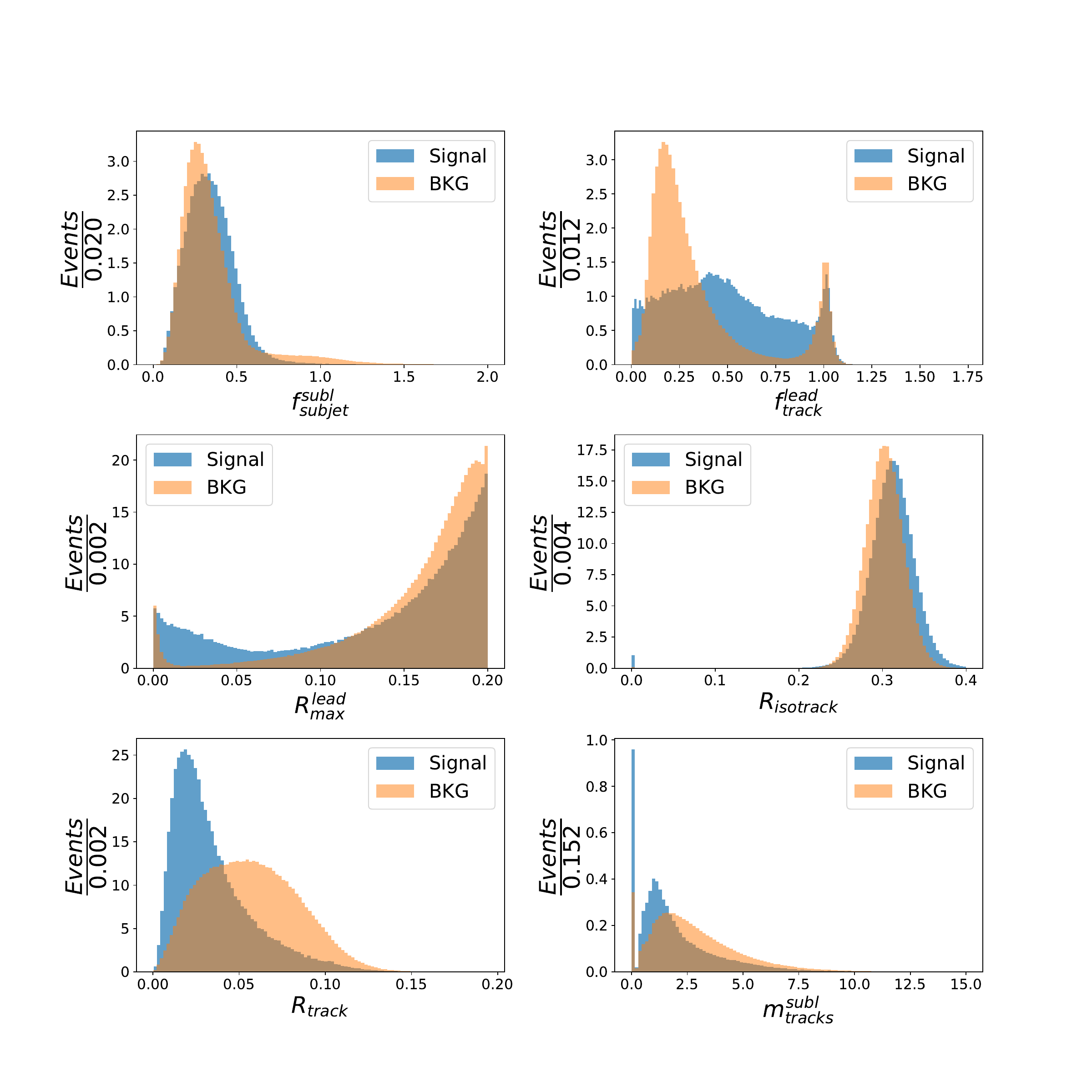}
        \caption{Unscaled BDT input features}
        \label{fig:unscaledfeats}
    \end{subfigure}
    \begin{subfigure}[ht]{0.35\paperwidth}
        \centering
        \includegraphics[width=0.35\paperwidth]{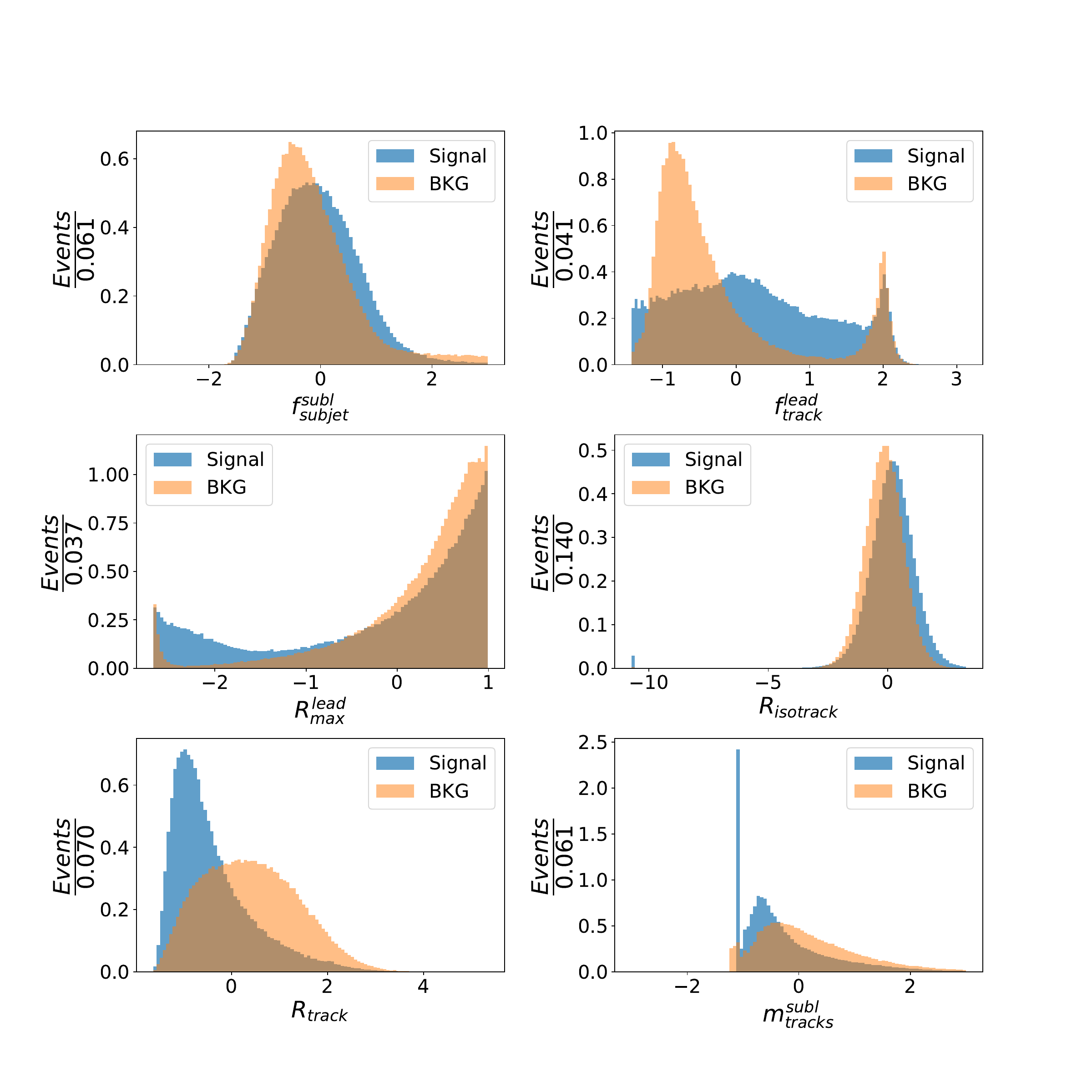}
        \caption{Standardized DNN input features}
        \label{fig:scaledfeats}
    \end{subfigure}
    \caption{High-level-feature distributions of di-$\tau$ objects in the sample dataset}
    \label{fig:inputfeats}
\end{figure}
While both the BDT and the DNN both use the same fixed-size set of input high-level-features per event, the DNN benefited greatly from standardization of the input variables, unlike the BDT which was far less sensitive to the absolute magnitudes of its input features. Standardization of input features guarantees that the distributions will have $\mu=0$ and $\sigma=1$. Distributions of selected input features before and after rescaling are given in Figure~\ref{fig:inputfeats}. 

For the CNN, the low-level-feature dataset needs to be encoded as an image. This is achieved by normalizing the position in the \(\eta\)-\(\phi\) plane with respect to the LRJ's direction. Furthermore, the energy of each individual pixel was scaled by the energy of the corresponding LRJ, such that it represents the fraction of the LRJ's $E_T$.
For the DS, in principle, the data does not need any special encoding. However, for ease of operation on the set, we treat each di-$\tau$ as an edge-less heterogeneous graph with three different node types corresponding to the sub-jets, calorimeter deposits and tracks within the LRJ. All constituents' $\{\eta,\Phi\}$ coordinates are shifted such that they represent the position with respect to the LRJ direction, while their $E_T$ values are scaled such that they represent the fraction of the LRJ's $E_T$.

\begin{figure}[t] 
    \centering
    \includegraphics[width=1.0\linewidth]{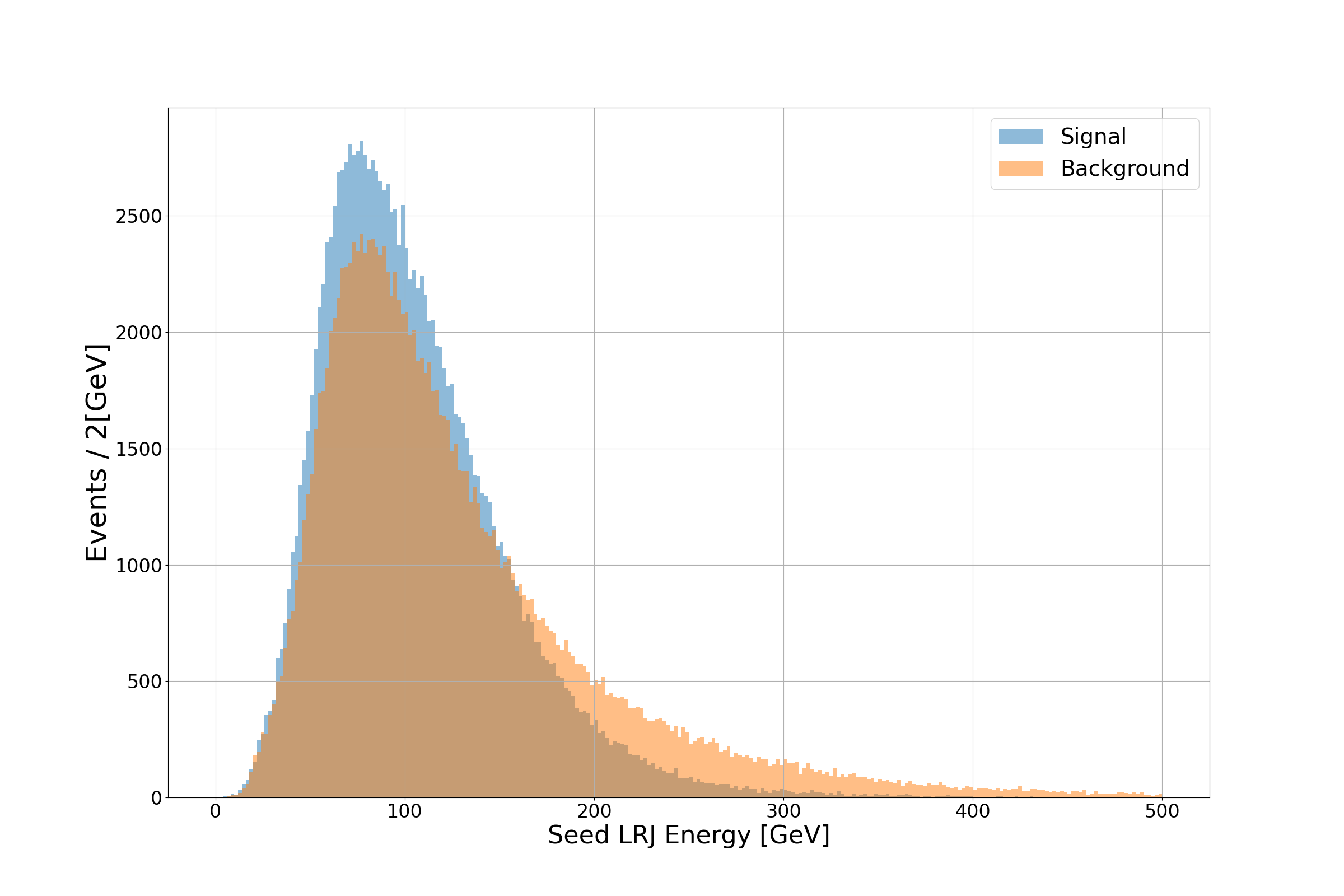}
    \caption{Distribution of LRJ transverse momentum in signal and background events. The LRJ's transverse momentum is used to scale its constituents' $E_T$, such that the NN learns to identify the di-$\tau$ based on relative-valued quantities rather than absolute ones.}
    \label{fig:LRJpt}
\end{figure}

\begin{figure}[h!]
    \centering
    \begin{subfigure}[ht]{0.35\paperwidth}
        \centering
        \includegraphics[width=0.35\paperwidth]{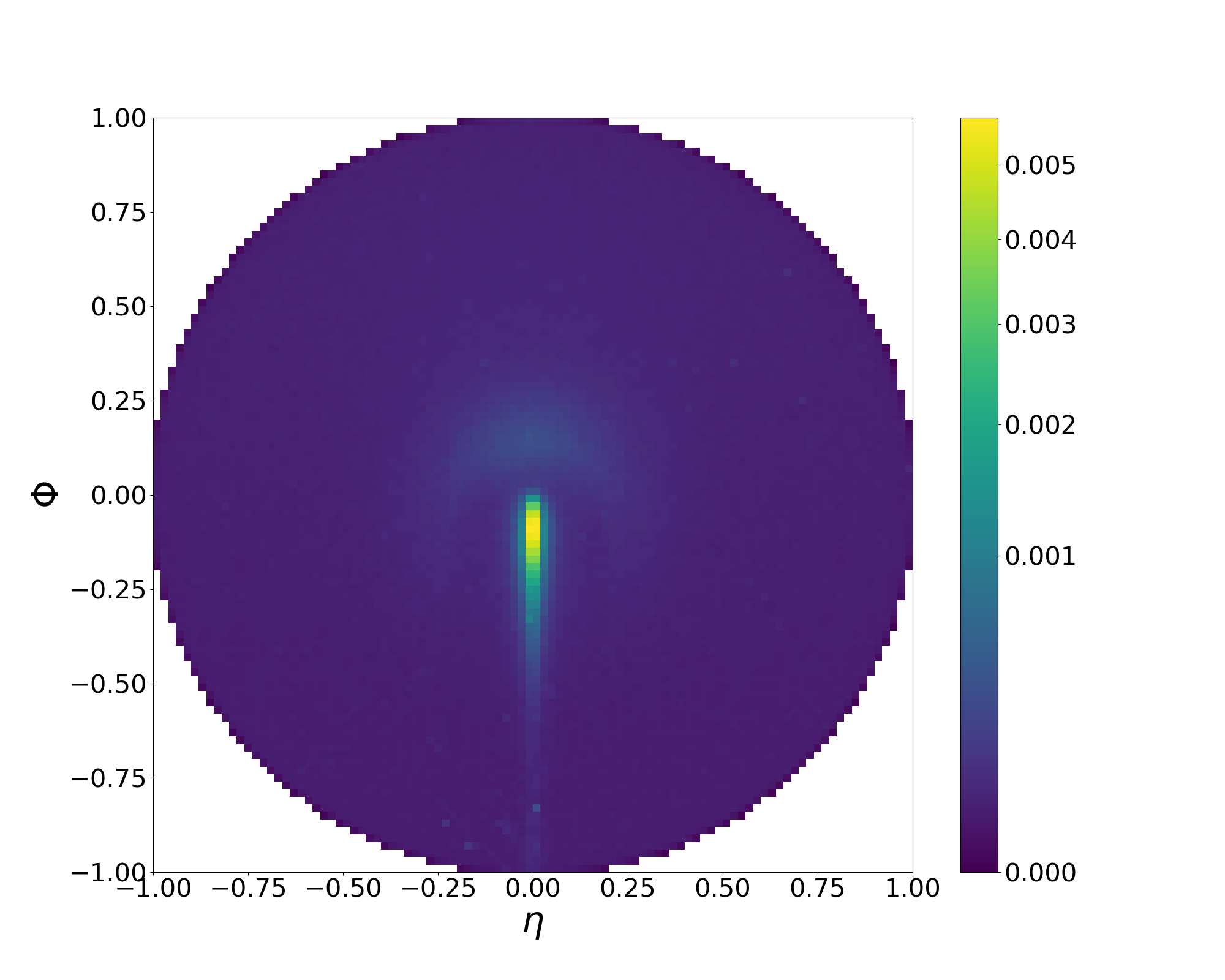}
        \caption{Signal (real) di-$\tau$}
        \label{fig:jetim_sig}
    \end{subfigure}
    \begin{subfigure}[ht]{0.35\paperwidth}
        \centering
        \includegraphics[width=0.35\paperwidth]{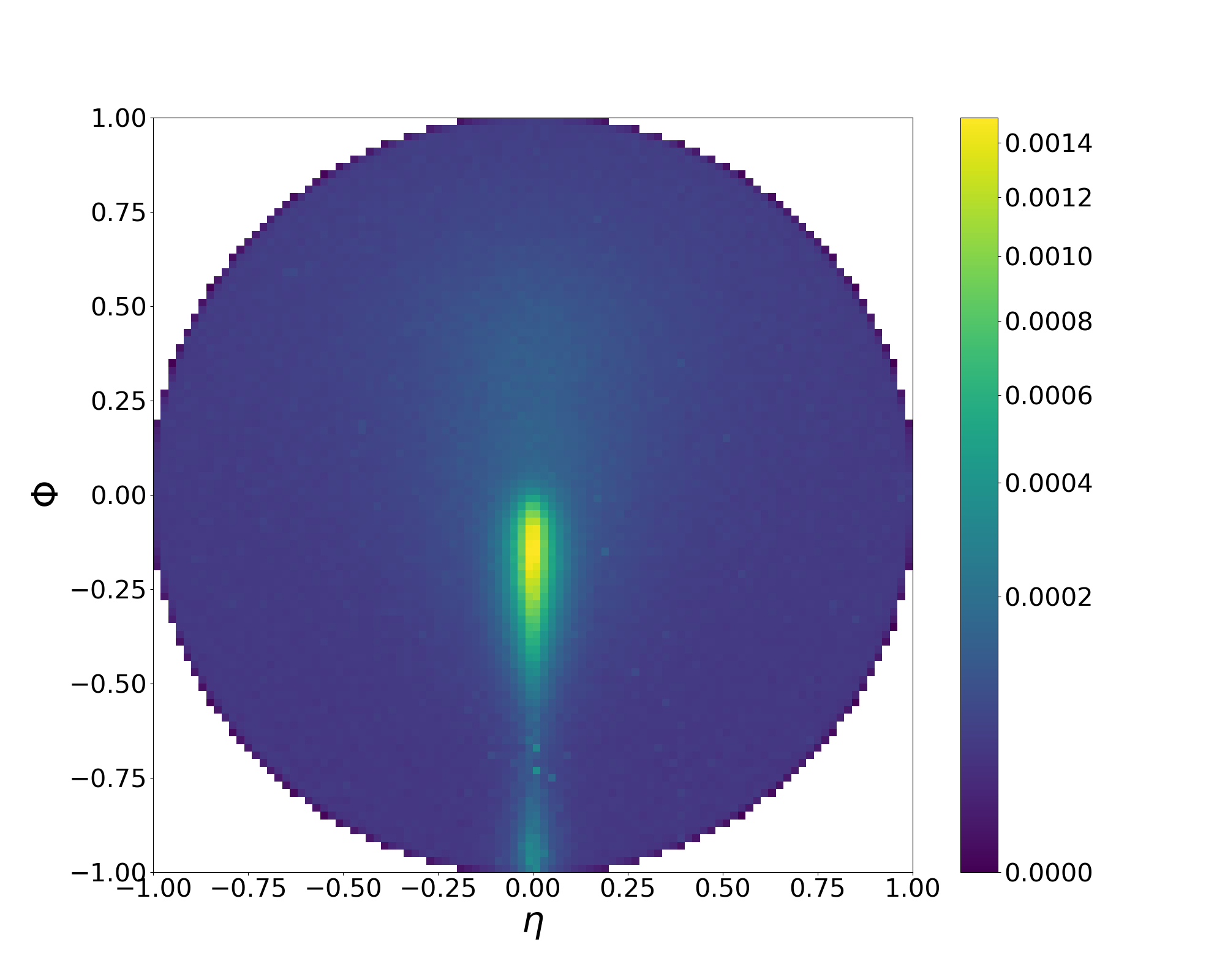}
        \caption{Background (fake) di-$\tau$}
        \label{fig:jetim_bkg}
    \end{subfigure}
    \caption{Distribution of scaled $E_T$ for calorimeter constituents of signal and background di-$\tau$ objects}
    \label{fig:jetimage}
\end{figure}

Figure~\ref{fig:LRJpt} shows the characteristic transverse momentum of signal (real) and background (fake) di-$\tau$ objects, by which the constituent $E_T$'s are scaled. An averaged distribution of scaled calorimeter constituent $E_T$ for signal and background di-$\tau$'s is given in Figure~\ref{fig:jetimage}. The constituents are rotated such that the leading sub-jet is centered on the negative y-axis, and their scaled transverse momenta are then collected into a normalized histogram. 

In the latter figure, the characteristically smaller angular spread of hard-energy deposits in real di-$\tau$'s manifests as both a faint but noticeable ring-shaped feature around the center of the LRJ originating from the sub-leading sub-jet, as well as a smaller average distance between the hard-energy deposits of the leading sub-jet aligned along the negative y-axis.

In total, the dataset consists of $\sim$400,000 events, evenly split between signal and background categories. It is split into training, validation and test sets with a 70\%-15\%-15\% ratio. The training and evaluation of the NNs is performed with PyTorch~\cite{pytorch}, which is also used for splitting the datasets into batches. The validation dataset is used during the NN's training to evaluate its performance, and the model with the lowest loss on the validation dataset is saved and kept as the nominal result of a training run.

\section{Results}
\begin{figure}[t] 
    \centering
    \includegraphics[width=1.0\linewidth]{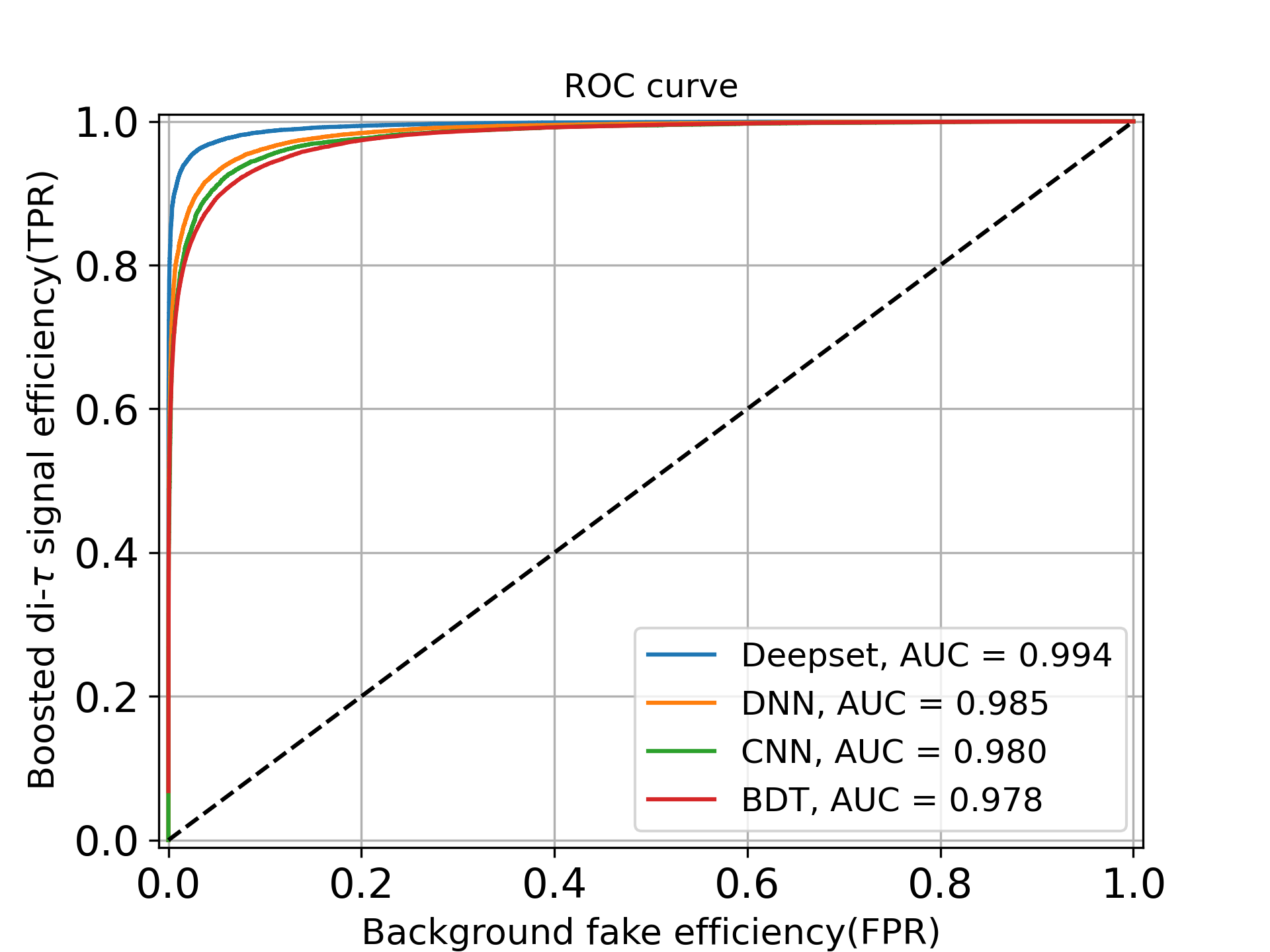}
    \caption{ROC curve,  signal efficiency (TPR) versus background efficiency (FPR), comparison among the models using high level information(BDT, DNN) and models using the low level information(CNN, Deepset). All neural network models are better than the baseline BDT model, and among them, Deepset performs the best. }
    \label{fig:results1}
\end{figure}

\raggedbottom
We compare the classification performance for the various models based on the receiver operating characteristic (ROC) curve and the area under the ROC curve (AUC) performance. The ROC curve is drawn using the di-$\tau$ signal efficiency, which is the true positive rate (TPR), versus background fake efficiency, which is the false positive rate (FPR). The ROC curves and their respective AUC's are shown in Figure~\ref{fig:results1}. The neural network models generally perform better than the baseline BDT model, and among them the DS shows the best rejection capabilities. The best DS model is obtained at the $77^\text{th}$ epoch, and achieves a loss value over the test set of 0.0962. Using a threshold value on network outputs of 0.5 for assigning class labels, this model achieves an accuracy over the test set of 96.34\%.

This can be attributed to several factors. The most crucial of these is that the DS includes the track impact parameter information as an input feature. Removal of this feature resulted in a significantly less powerful model, decreasing the ROC-AUC to 0.981 - and is thus likely related to its outperforming the CNN, which also uses only the low-level variables. In a similar context, the DS also considers the sub-jet kinematic information, which isn't used in the CNN, and further strengthens its predictive ability. As previously mentioned, the DS is able to encode global information efficiently into its latent representations, which also improves its classification performance - a DS model without these global aggregation steps has also been tested and required a factor of $\sim\times3$ more parameters to achieve a similar ROC-AUC. Lastly, the attention-pooling layers allow the DS to represent the physics-motivated concept that certain constituents of the di-$\tau$ are more important than others, thus introducing an inductive bias via a learned-weight aggregation of the different set elements before passing them to the MLP classification head.

The CNN demonstrates better discriminating power than the BDT, but worse than the other models tested. The best CNN model is obtained at the $14^\text{th}$ epoch, and achieves a loss value over the test set of 0.1793. Using a threshold value on network outputs of 0.5 for assigning class labels, this model achieves an accuracy over the test set of 93.07\%.  This may be attributed to the simplified detector simulation used in these studies, which includes only a single layer in the ECAL and HCAL, and thus does not allow for a more complicated shower shape description. In addition, particles in a jet often deposit energy in relatively few cells, and as a result jet images tend to be fairly sparse, weakening the inherent power of CNNs to achieve good performance with relatively few parameters. It is also worth mentioning that, since jet images are rather different than common pictures on which CNNs have well-demonstrated potential, the use of pre-trained kernels (i.e. "transfer-learning") is less applicable to the problem of jet classification.

Of the NN-based models, the DNN is the simplest one examined. After feature standardization, the DNN consistently performs better than the BDT, while using a very basic architecture. The best DNN model is obtained at the $80^\text{th}$ epoch, and achieves a loss value over the test set of 0.1544. Using a threshold value on network outputs of 0.5 for assigning class labels, this model achieves an accuracy over the test set of 93.98\%. This demonstrates that introduction of non-linearities via the activation functions allow the model to take better advantage of the discriminating power inherent in the high-level features, and as such remains the most interpretable out of the NN-based models tested.

While the DS and DNN models reported in this paper use a similar number of learnable parameters ($\sim$17,000), this is not the case for either the BDT or CNN models, which use approximately a factor of three and four more parameters, respectively. As the BDT and CNN demonstrate an overall inferior performance compared to the DS and DNN, a larger number of parameters is chosen to better explore the potential discrimination power of these models. In the case of the DNN, increasing the number of parameters further does not result in any significant performance improvement - and though for the DS an improved performance can indeed be obtained with more parameters, the smaller-sized DS is chosen to provide a more substantiated comparison with the DNN as the two best-performing architectures. Smaller variants of the BDT and CNN, using a similar number of parameters as the DS/DNN, yield ROC AUC's of approximately 0.944 and 0.970, respectively - lending further credence to the observation of the architectures' performance hierarchy, which in this case becomes more distinct with much larger differences between the four studied model types.
\section{Conclusion}
A comparison of four different models for classification of boosted di-$\tau$ topologies was performed, encompassing two different paradigms -- using low-level constituent information and high-level custom variables. The models were intentionally kept simple, as the goal was not to develop a novel method but rather compare the performance between the two different paradigms, as well as examine whether NN-based models outperform a BDT-based model. The latter was confirmed to be true, while the former was found to depend heavily on use of particular low-levels features with strong discriminating power. Realistically, however, a CNN may benefit from a more detailed calorimeter simulation than was used in these studies, and a DS can easily be extended to accomodate both high-level and low-level features via replication and concatenation of the high-level variables to the low-level constituents. We conclude that an ideal classification model will thus include both complementary types of information, but this is left for future works.

\appendix
\section{PCA Decomposition}
\label{app:PCA}
\begin{figure}[h!]
    \centering
    \begin{subfigure}[h!b]{0.35\paperwidth}
        \centering
        \includegraphics[width=0.35\paperwidth]{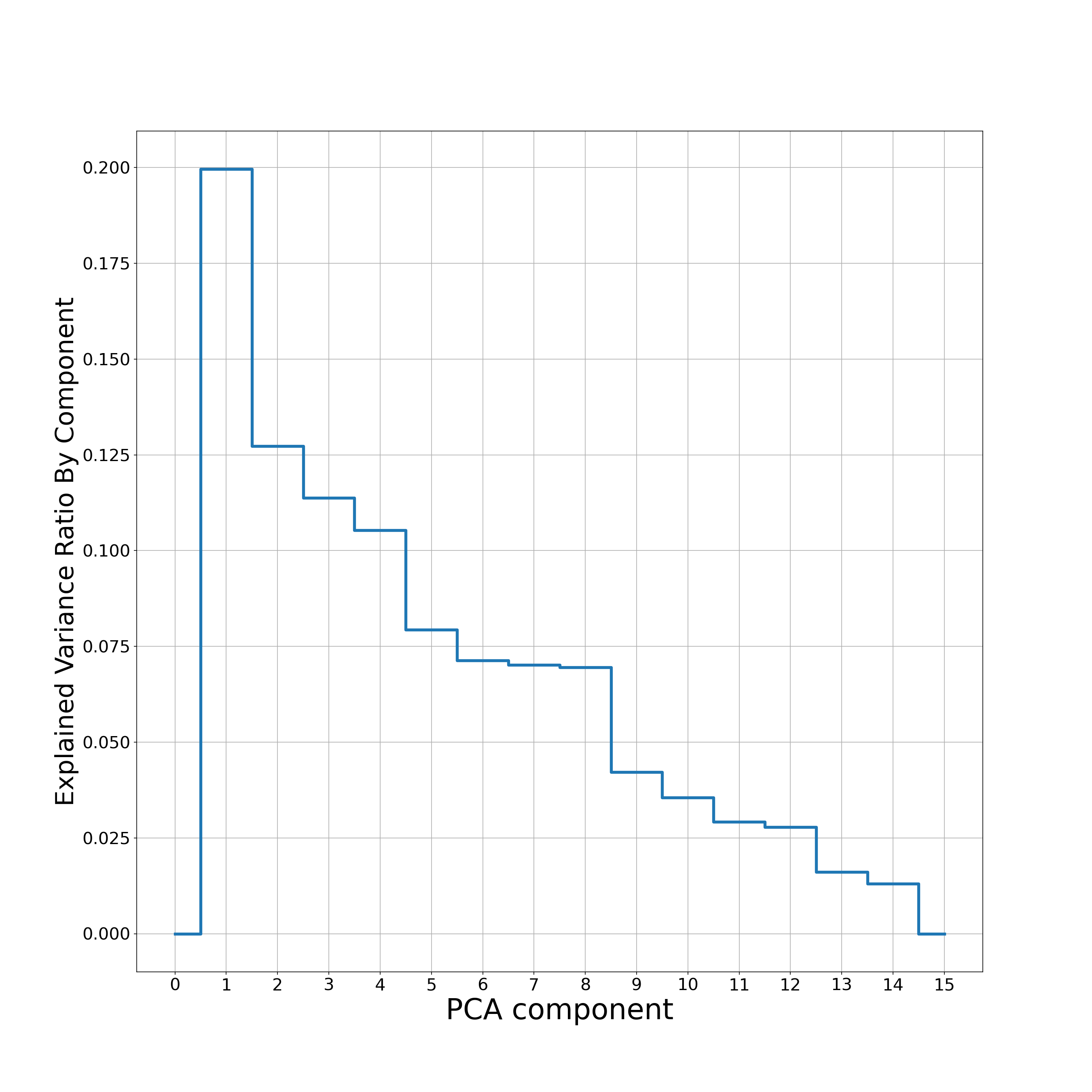}
        \caption{\centering{Fraction of total variance explained by each principal component}}
    \end{subfigure}
    \begin{subfigure}[h!b]{0.35\paperwidth}
        \centering
        \includegraphics[width=0.35\paperwidth]{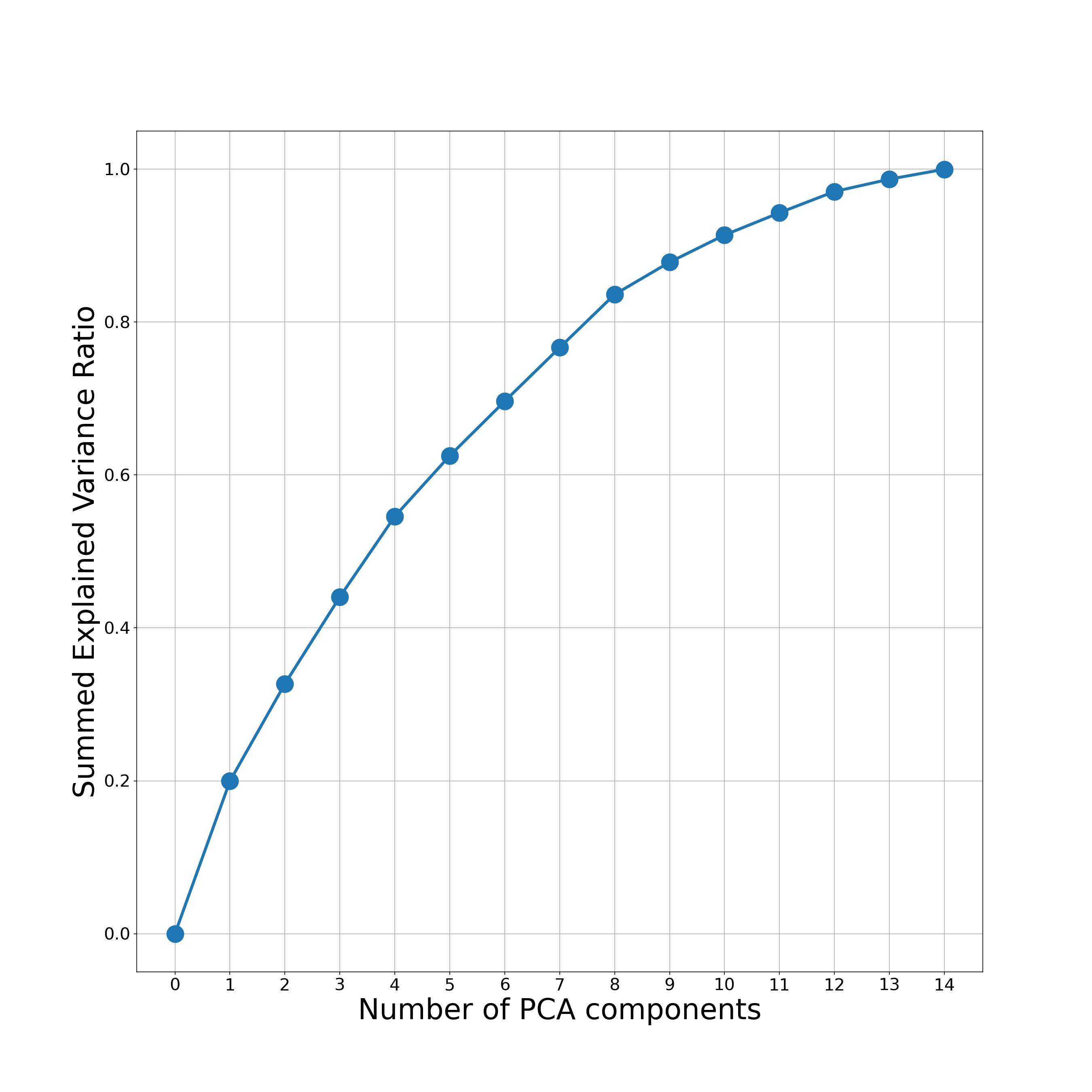}
        \caption{\centering{Summed fraction of total variance explained by the leading N principal components}}
    \end{subfigure}
    \caption{Explained variance ratios extracted from PCA decomposition of input feature set.}
    \label{fig:PCAvar}
\end{figure}

As mentioned, a PCA decomposition study was conducted on the high-level features in the dataset. 
The dependence of the explained variance ratio on the number of principal components is illustrated in Figure~\ref{fig:PCAvar}, showing that no principal component captures a dominant fraction of the total variance - perhaps other than the first principal component, which accounts for approximately 20\% of the variance - and that adding components results in a fairly consistent, slow increase in the fraction of variance captured.

\begin{figure}[h!] 
    \centering
    \includegraphics[width=1.0\linewidth]{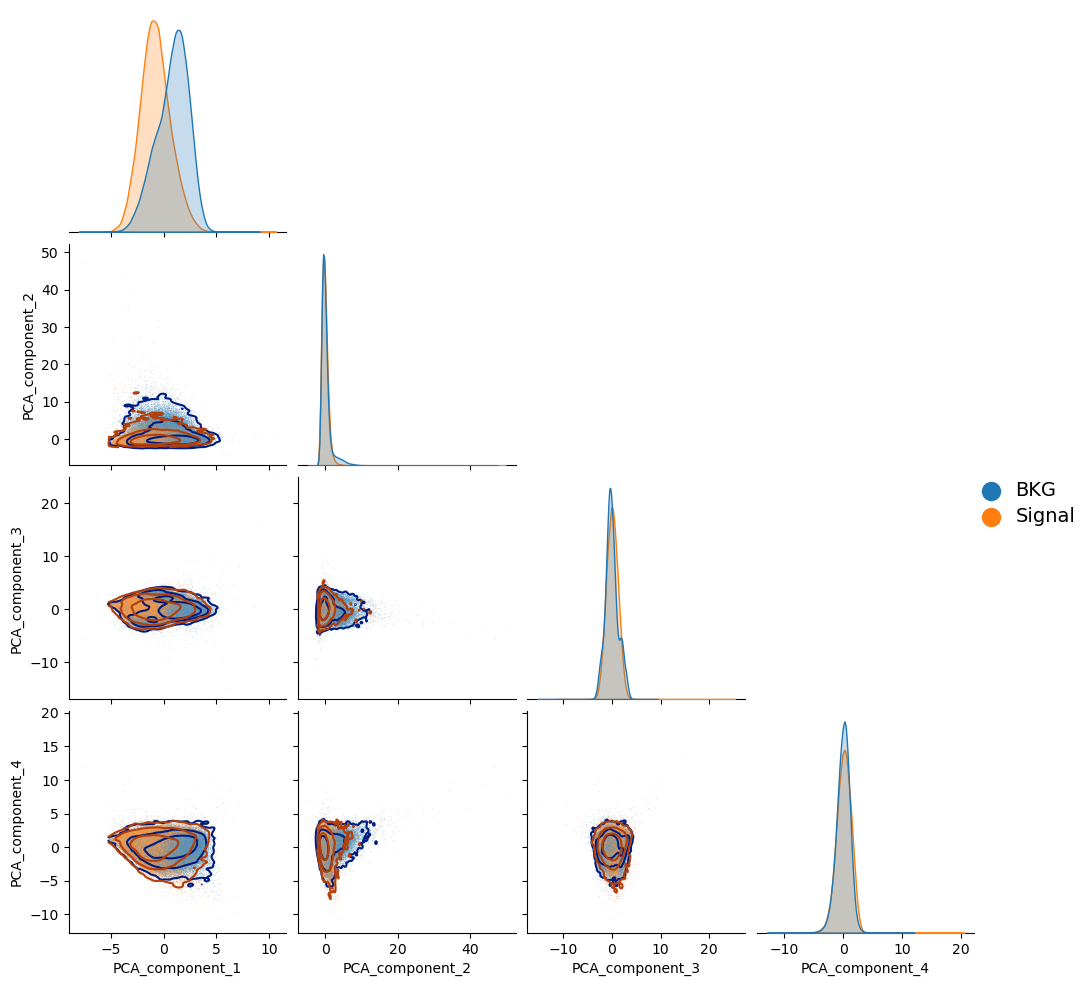}
    \caption{Distribution of each of the four leading principal components versus the other three.}
    \label{fig:PCAcorner}
\end{figure}

\begin{figure}[h!]
    \centering
    \begin{subfigure}[h!]{1.05\linewidth}
        \centering
        \includegraphics[width=1.05\linewidth]{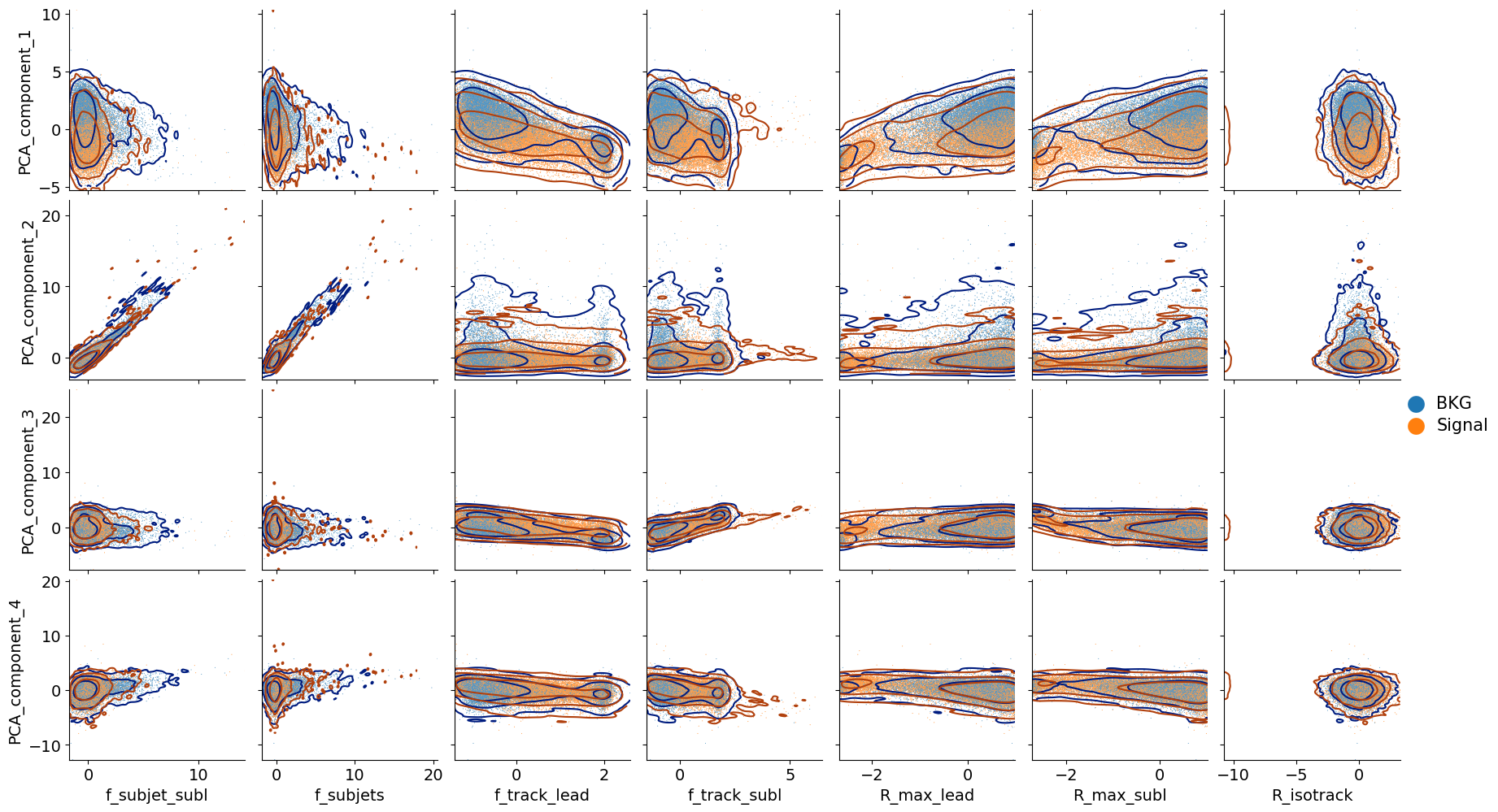}
    \end{subfigure}
    \\[5ex]
    \begin{subfigure}[h!]{1.05\linewidth}
        \centering
        \includegraphics[width=1.05\linewidth]{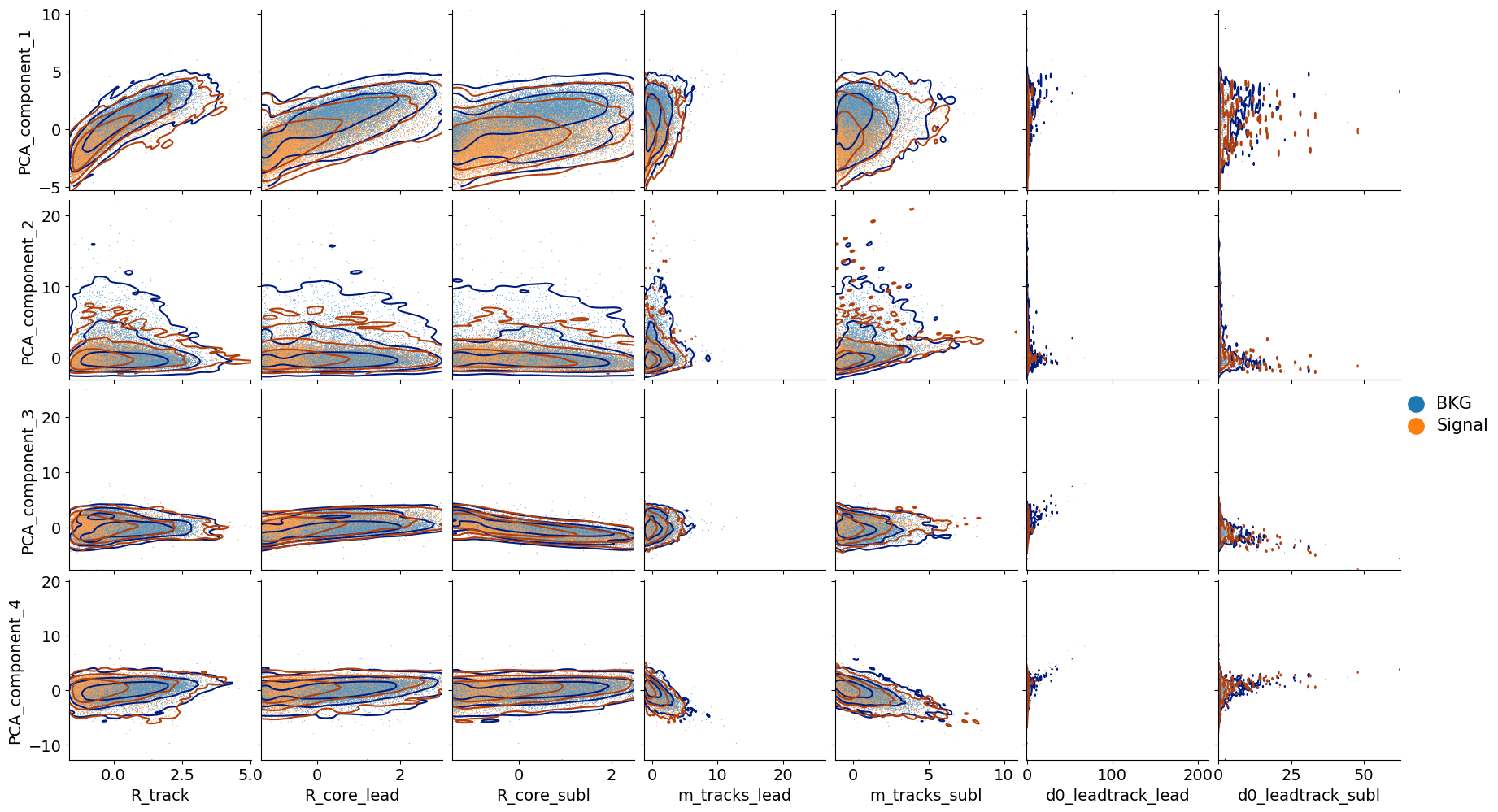}
    \end{subfigure}
    \caption{Distributions of the four leading principal components versus the high-level input features used in the BDT and DNN.}
    \label{fig:PCAvsInputs}
\end{figure}

For the leading four principal components, Figure~\ref{fig:PCAcorner} shows each component's behavior versus the other three (along with the component's distribution given on the diagonal), while Figure~\ref{fig:PCAvsInputs} shows each component's behavior as a function of the fourteen input variables along with kernel density estimate ("KDE") contours corresponding to a coverage of 1-, 2- and 3- standard deviations. The KDE contours are placed in order to assist the visualization and interpretation of the joint probability distributions. 
From these latter figures, it is evident that the 2nd principal component ("sub-leading") is very highly correlated with two of the variables ($f_{subjet}^{sublead}$ and $f_{subjets}$, which are known to be correlated among themselves based on the definition of the features). No particular striking clustering patterns emerge for the different classes (in some cases there is some soft clustering, mostly involving PC2 which as mentioned correlates very strongly with two of the features), illustrating that indeed given these variables the problem is difficult to solve purely with linear transformations.\newpage ~ \vfill
\vfill
\newpage
\newpage
\section{Manifold learning decompositions}
\label{app:Manifold}
For completeness' sake, we report here the results of decompositions derived using t-SNE~\cite{tSNE_paper} and UMAP~\cite{umap_paper} methods on the high-level features in the dataset. The decomposition is performed with the openTSNE~\cite{tSNE_python} and UMAP~\cite{umap_python} python packages, using a PCA-initialization and a euclidean distiance metric. These methods explore non-linear relations, often revealing clustering patterns present in the data which do not appear in the PCA decomposition. In our implementation, the t-SNE and UMAP hyperparameters balancing local and global structure were optimized manually to yield better and more meaningful component distributions. 

\begin{figure}[h!] 
    \centering
    \includegraphics[width=1.0\linewidth,height=0.4\textheight,keepaspectratio=true]{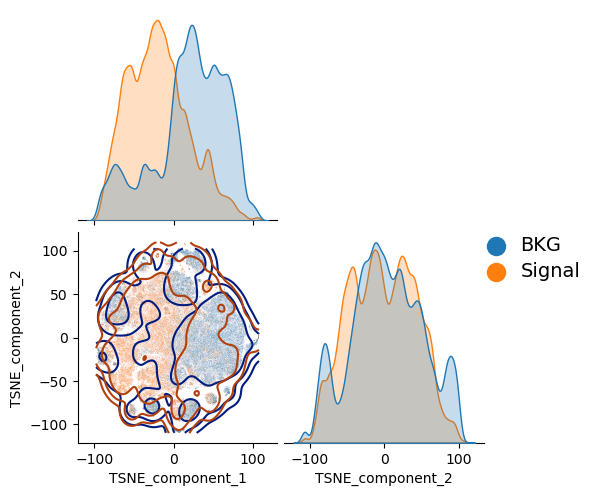}
    \caption{Joint distribution of the fitted t-SNE components.}
    \label{fig:TSNEcorner}
\end{figure}

\begin{figure}[h!] 
    \centering
    \includegraphics[width=1.0\linewidth,height=0.4\textheight,keepaspectratio=true]{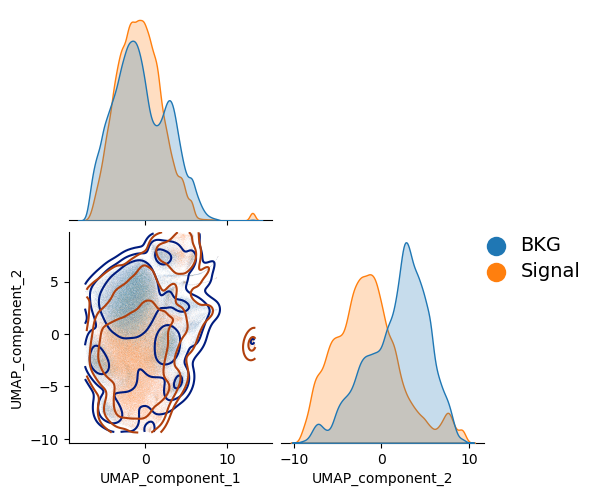}
    \caption{Joint distribution of the fitted UMAP components.}
    \label{fig:UMAPcorner}
\end{figure}

\begin{figure}[h!]
    \centering
    \begin{subfigure}[h!]{1.05\linewidth}
        \centering
        \includegraphics[width=1.05\linewidth,height=0.15\textheight,keepaspectratio=true]{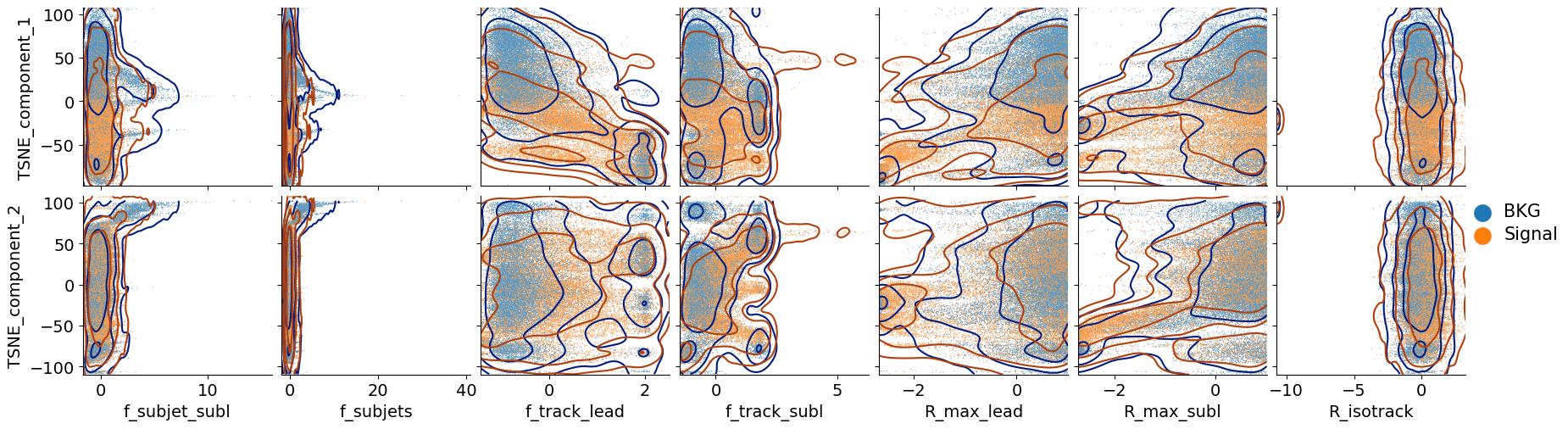}
    \end{subfigure}
    \\[5ex]
    \begin{subfigure}[h!]{1.05\linewidth}
        \centering
        \includegraphics[width=1.05\linewidth,height=0.15\textheight,keepaspectratio=true]{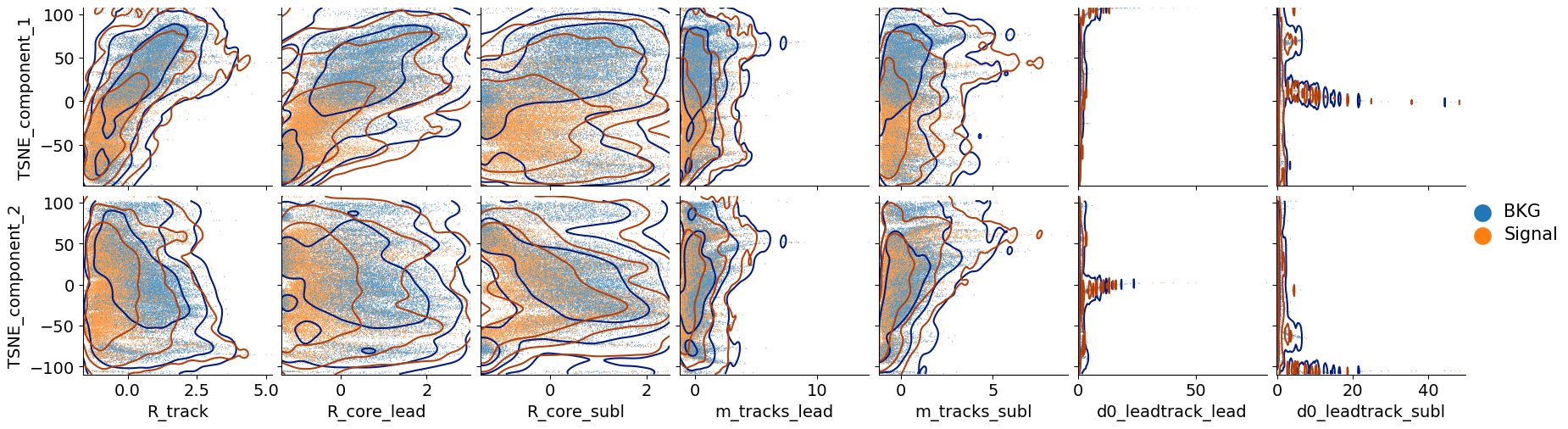}
    \end{subfigure}
    \caption{Distributions of the fitted t-SNE components versus the high-level input features used in the BDT and DNN.}
    \label{fig:TSNEvsInputs}
\end{figure}

\begin{figure}[h!]
    \centering
    \begin{subfigure}[h!]{1.05\linewidth}
        \centering
        \includegraphics[width=1.05\linewidth,height=0.15\textheight,keepaspectratio=true]{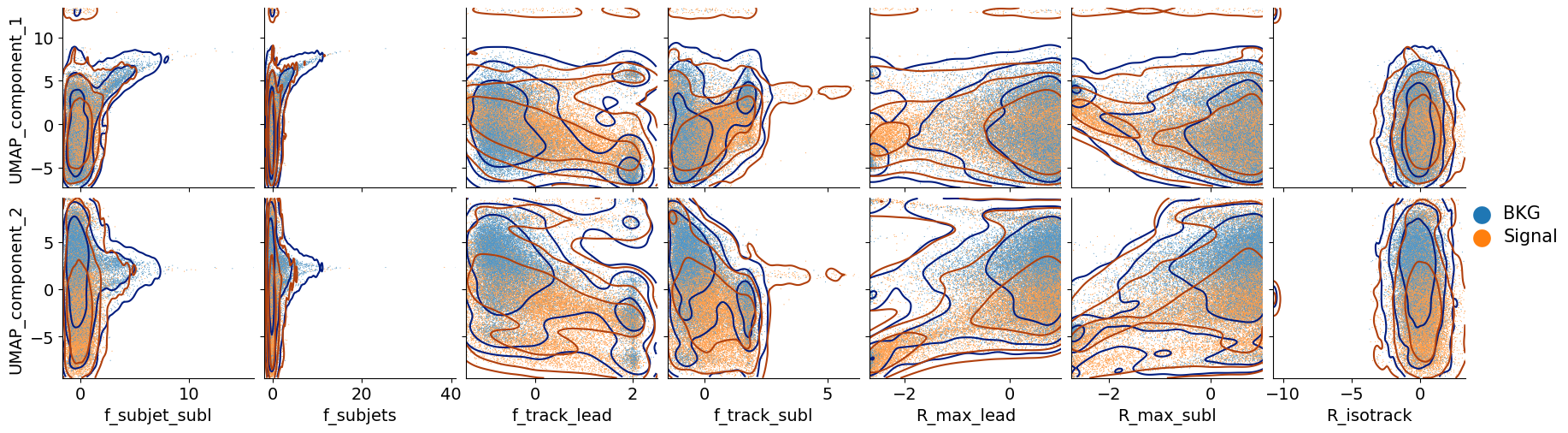}
    \end{subfigure}
    \\[5ex]
    \begin{subfigure}[h!]{1.05\linewidth}
        \centering
        \includegraphics[width=1.05\linewidth,height=0.15\textheight,keepaspectratio=true]{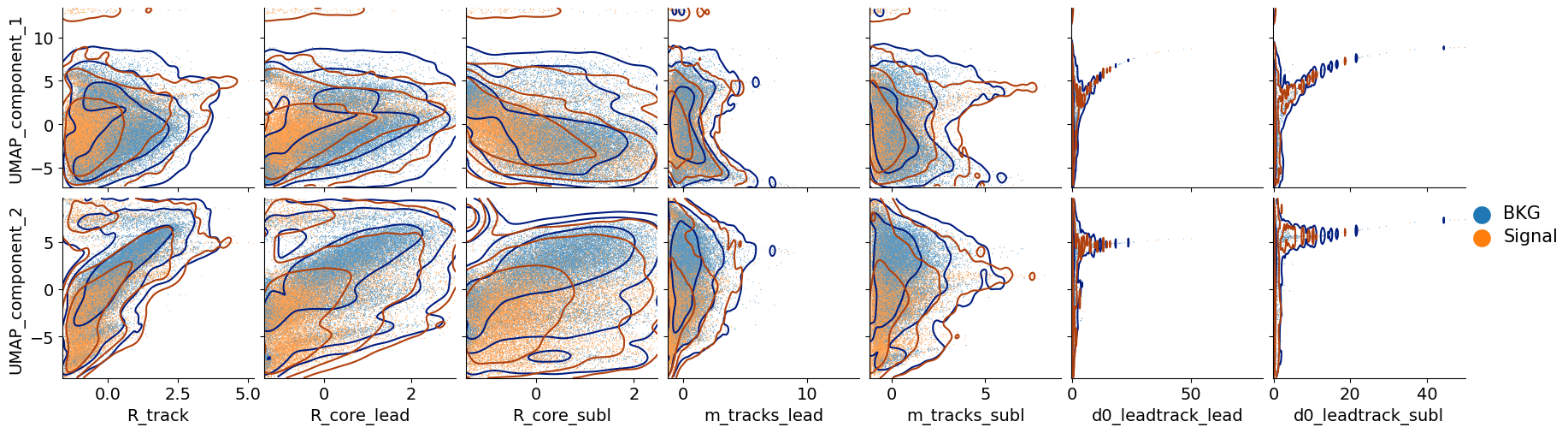}
    \end{subfigure}
    \caption{Distributions of the fitted UMAP components versus the high-level input features used in the BDT and DNN.}
    \label{fig:UMAPvsInputs}
\end{figure}

For each method, two components are extracted, whose joint distribution is given in Figure~\ref{fig:TSNEcorner} for t-SNE and Figure~\ref{fig:UMAPcorner} for UMAP (along with the component's distribution given on the diagonal). Figures~\ref{fig:TSNEvsInputs} and~\ref{fig:UMAPvsInputs} show each component's behavior as a function of the fourteen input variables for the t-SNE and UMAP decompositions, respectively. Kernel density estimate ("KDE") contours corresponding to a coverage of 1-, 2- and 3- standard deviations are placed in order to assist the visualization and interpretation of the joint probability distributions.

As these plots illustrate, unlike the straightforward PCA, the manifold-learning methods both manage to cluster the two classes at least partially, likely due to different underlying nonlinear relations present between the high-level features in each class. Of the two, the t-SNE seems to generate slightly more distinct clusters, displaying finer structure within each class. Furthermore, both methods yield a single component with greater discriminating power than the first principal component, though neither method on its own is competitive in terms of performance with the algorithmns studied in the body of this paper. 

However, when compared to the PCA study, the increased performance of the manifold learning methods supports the notion of nonlinear relations between the variables being of key importance for designing a successful algorithm for boosted di-$\tau$ classification.

\newpage ~ \vfill
\vfill
\newpage
\newpage
\section{BDT and NNs}
\label{app:BDTNN}
\subsection{Boosted decision tree}
The BDT model is trained with TMVA~\cite{TMVA} package. A total of 14 variables, defined in Section~\ref{highlevel}, are used as the input features. It is trained using the Adaptive Boosting and setting of: $\text{NTrees}=1000$, $\text{MaxDepth}=3$, $\text{AdaBoostBeta}=0.2$, $\text{BaggedSampleFraction}=0.3$, $\text{SeparationType}=\text{GiniIndex}$, \text{nCuts}=300$, $\text{MinNodeSize}=0.5\%. The total number of parameters can be estimated as the number of decision nodes times the number of variables, resulting in 49420 "trainable parameters".

\subsection{Deep neural network}
The deep neural network(DNN) uses the same input features as the BDT. A user-defined latent representation of dimension $\mathcal{D}^h$ is used for the internal layers ($\mathcal{D}^h=48$ in our implementation). The initialization head uses a $14\rightarrow32\rightarrow \mathcal{D}^h$ structure. It is followed up by five $\mathcal{D}^h\rightarrow \mathcal{D}^h$ layers, and then a classification head using a $\mathcal{D}^h\rightarrow32\rightarrow16\rightarrow1$ reduction structure. A 10\% dropout layer follows each activation function. Including the learned parameters in the PReLU activation functions, the total number of parameters in the DNN used in this paper is 15946. The model was trained with a batch size of 2038, a learning rate of $7.8\times10^{-4}$, L2 normalization weight of $4.9\times10^{-5}$ and a dropout fraction of $8.8$\%, where the values were obtained from an optimization study using Optuna.
\subsection{Convolutional neural network}
The network takes an input image of the simulated data from the calorimeters and the tracker and processes it through various layers. These layers are organized as three convolutional blocks, followed by a data flattener, and ending with a classifier block. Each convolutional block is made of a convolutional layer that takes $N_\text{in}$ input channels and returns $N_\text{out}$ output channels, and has a squared kernel; a Parametrized ReLU layer; a 10\% Dropout layer; and a 2D Max Pooling layer with a square kernel of size 2 and a stride of 2. The first and second convolutional kernels have a size of 5, whereas the third one has a size of 3. The first convolutional block takes 3 input channels and returns 16 output channels, the second block takes 16 and returns 32, and the third takes 32 and returns 64. Following the data flattener, the classifier block is made of two linear subblocks made of a linear layer, a Parametrized ReLU layer, and a Dropout layer; and after these two blocks comes one more linear layer. The first layer takes the flattened data (of size 1600) and outputs into 32 nodes, the second layer outputs into 16 nodes, and the last layer outputs into one node giving the final output. Including the learned parameters in the PReLU activation functions, the total number of parameters in the CNN used in this paper is 84326.

\subsection{DeepSet neural network}
The structure of the DS implementation is detailed in the body of the paper. The input (latent) representation dimensions for input type $j$ are denoted as $\mathcal{D}^{in}_j$ ($\mathcal{D}^h_j$). The initialization NN's use a $\mathcal{D}^{in}_j\rightarrow\mathcal{D}^h_j\rightarrow\mathcal{D}^h_j$ structure. The classification head uses a $\sum_j\mathcal{D}^h_j\rightarrow\sum_j\mathcal{D}^h_j\rightarrow\sum_j\mathcal{D}^h_j\rightarrow32\rightarrow16\rightarrow1$ structure. Each of the three DS blocks for each input type uses a $\left[(4\times\mathcal{D}^h_j)+\mathcal{D}^{in}_j\right]\rightarrow\mathcal{D}^h_j\rightarrow\mathcal{D}^h_j\rightarrow\mathcal{D}^h_j$ structure. A 10\% dropout layer follows each activation function. The latent representation dimensions chosen are $\mathcal{D}^h_j=4\times\mathcal{D}^{in}_j$. Including the learned parameters in the PReLU activation functions, the total number of free parameters in the DS used for this paper is 17652.  The model was trained with a batch size of 808, a learning rate of $2.6\times10^{-3}$, L2 normalization weight of $6.2\times10^{-5}$ and a dropout fraction of $5.9$\%, where the values were obtained from an optimization study using Optuna.

\section{Training and testing history and results}

\subsection{BDT}
\begin{figure}[h]
    \centering
    \includegraphics[width=\linewidth,height=0.3\textheight,keepaspectratio=true]{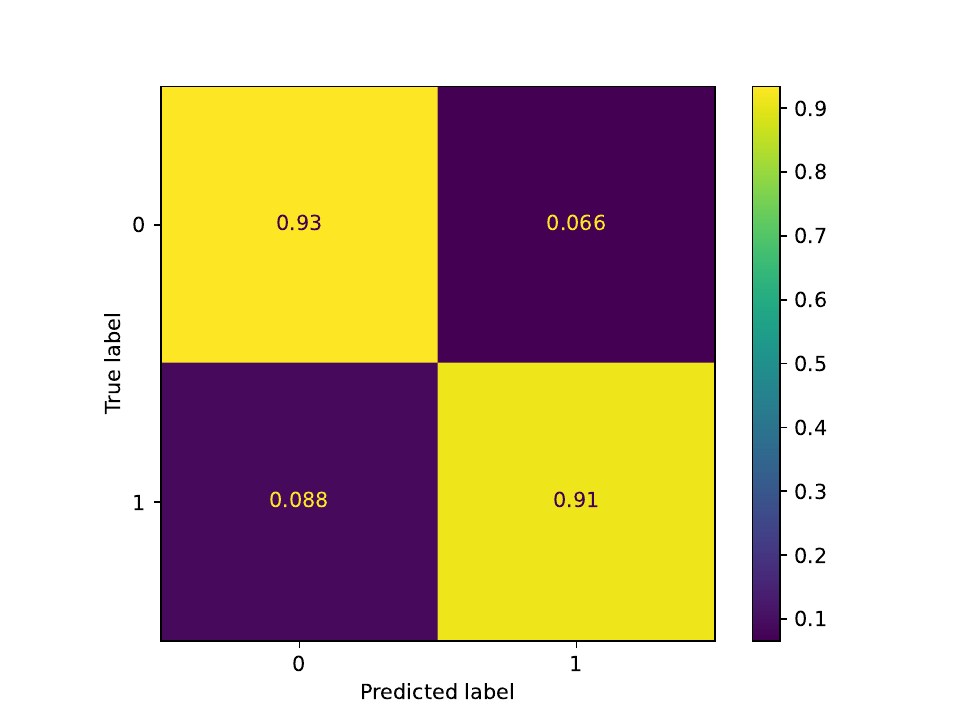}
    \caption{Confusion matrix for the BDT testing with a threshold of 0. The resulting accuracy value (defined as the rate of total correct predictions) is 93.07\%. The resulting precision value (defined as the ratio between the number of true signal predictions and the number of total signal predictions) is 72.45\%.}
\end{figure}
\newpage

\subsection{DNN}

\begin{figure}[h]
    \centering
    \includegraphics[width=\linewidth,height=0.35\textheight,keepaspectratio=true]{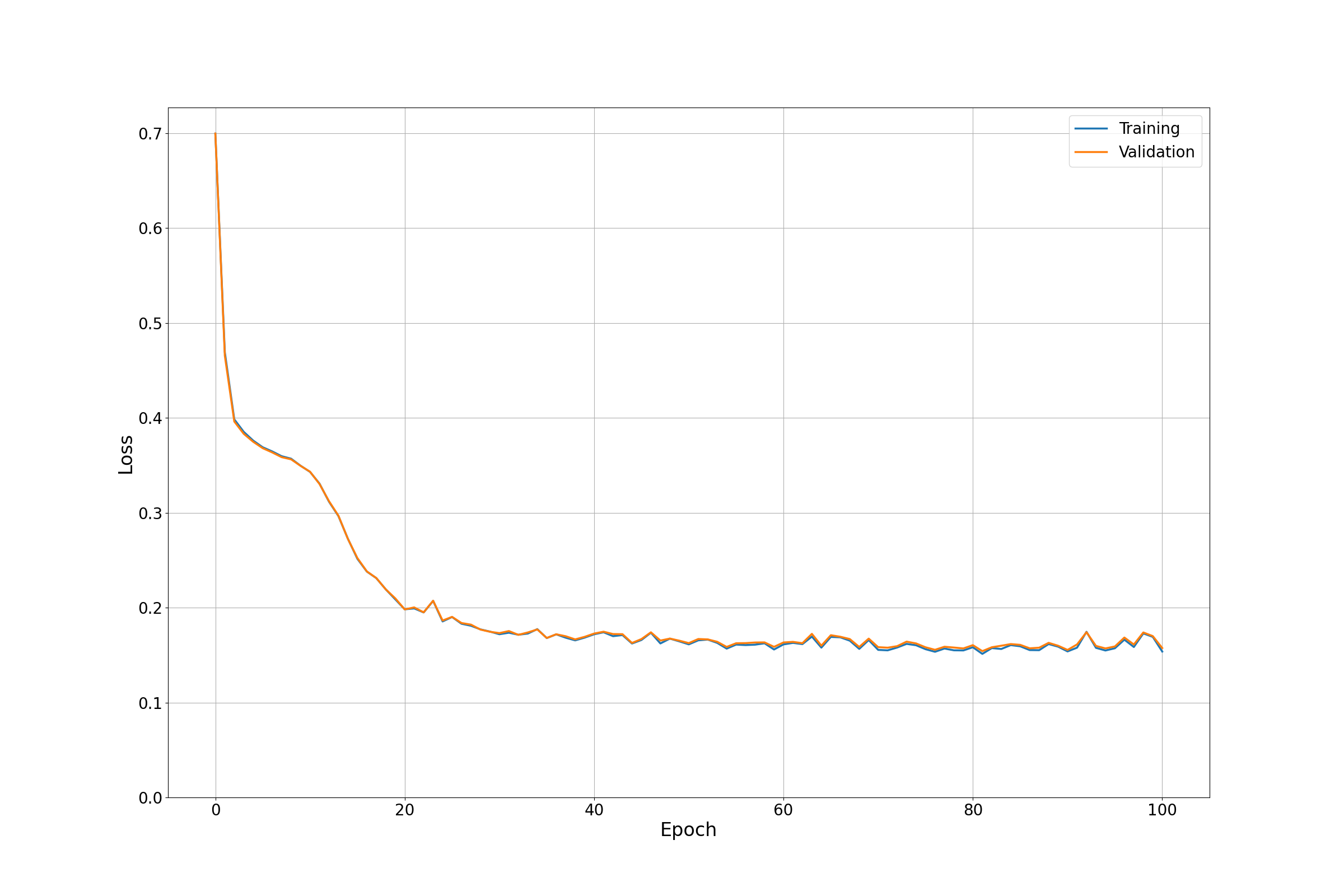}
    \caption{Loss plots of DNN training and validation at each epoch for the selected DNN model. The minimal loss is at epoch 80. The selected model yields a testing loss of 0.15436.}
\end{figure}
\begin{figure}[h]
    \centering
    \includegraphics[width=\linewidth,height=0.3\textheight,keepaspectratio=true]{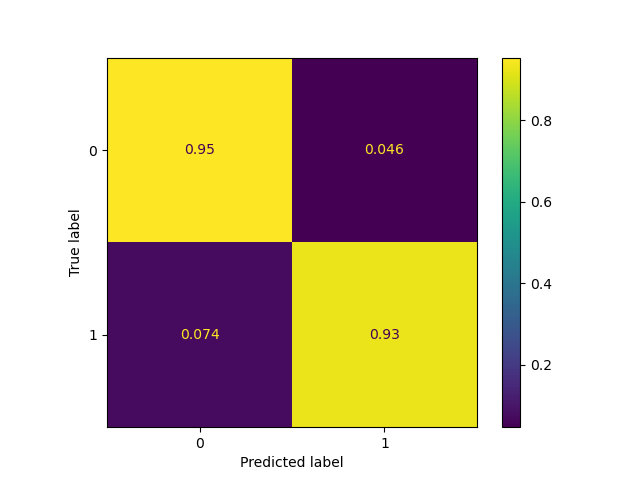}
    \caption{Confusion matrix for the DNN testing with a threshold of 0.5. The resulting accuracy value (defined as the rate of total correct predictions) is 93.98\%. The resulting precision value (defined as the ratio between the number of true signal predictions and the number of total signal predictions) is 95.21\%.}
\end{figure}
\newpage
\subsection{CNN}

\begin{figure}[h]
    \centering
    \includegraphics[width=\linewidth,height=0.35\textheight,keepaspectratio=true]{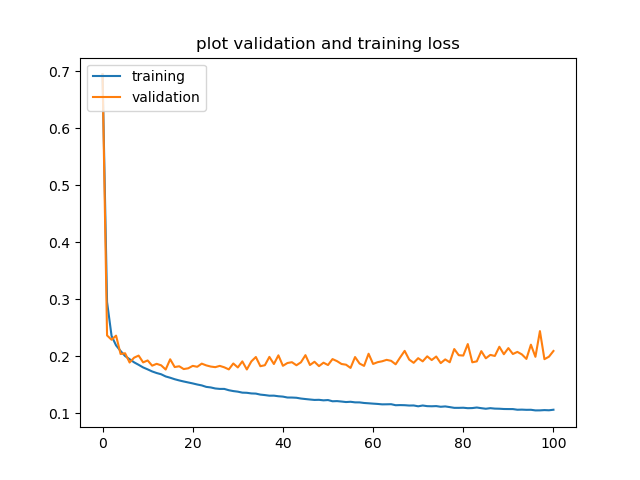}
    \caption{Loss plots of CNN training and validation at each epoch for the selected CNN model. The minimal loss is at epoch 14. The selected model yields a testing loss of 0.17926.}
\end{figure}

\begin{figure}[h]
    \centering
    \includegraphics[width=\linewidth,height=0.3\textheight,keepaspectratio=true]{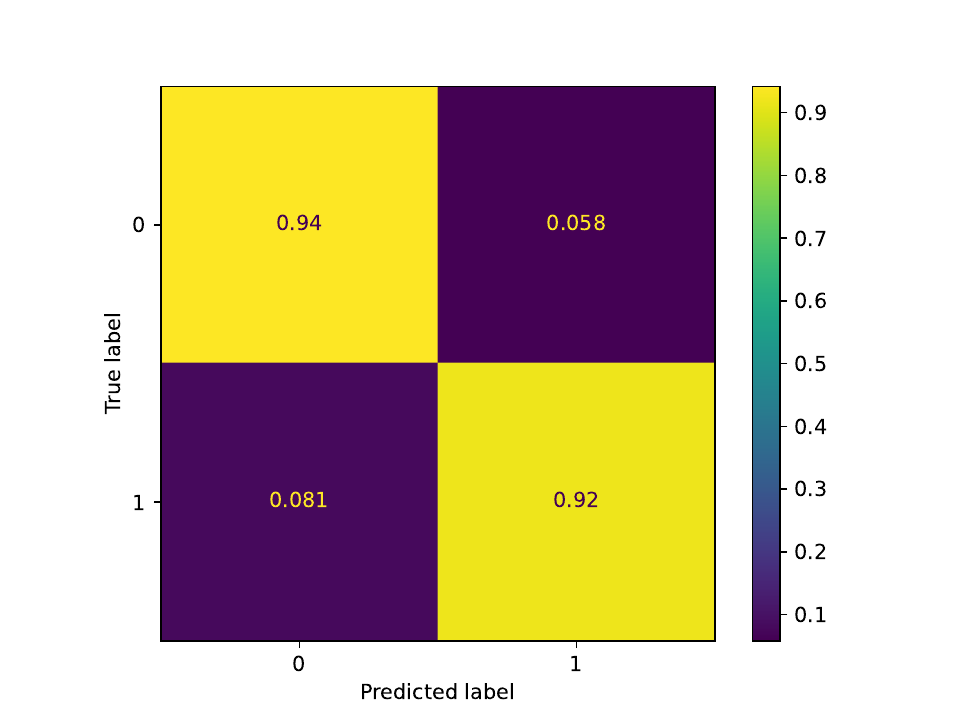}
    \caption{Confusion matrix for the CNN testing with a threshold of 0.5. The resulting accuracy value (defined as the rate of total correct predictions) is 93.07\%. The resulting precision value (defined as the ratio between the number of true signal predictions and the number of total signal predictions) is 93.91\%.}
\end{figure}
\newpage
\subsection{DeepSet}

\begin{figure}[h]
    \centering
    \includegraphics[width=\linewidth,height=0.35\textheight,keepaspectratio=true]{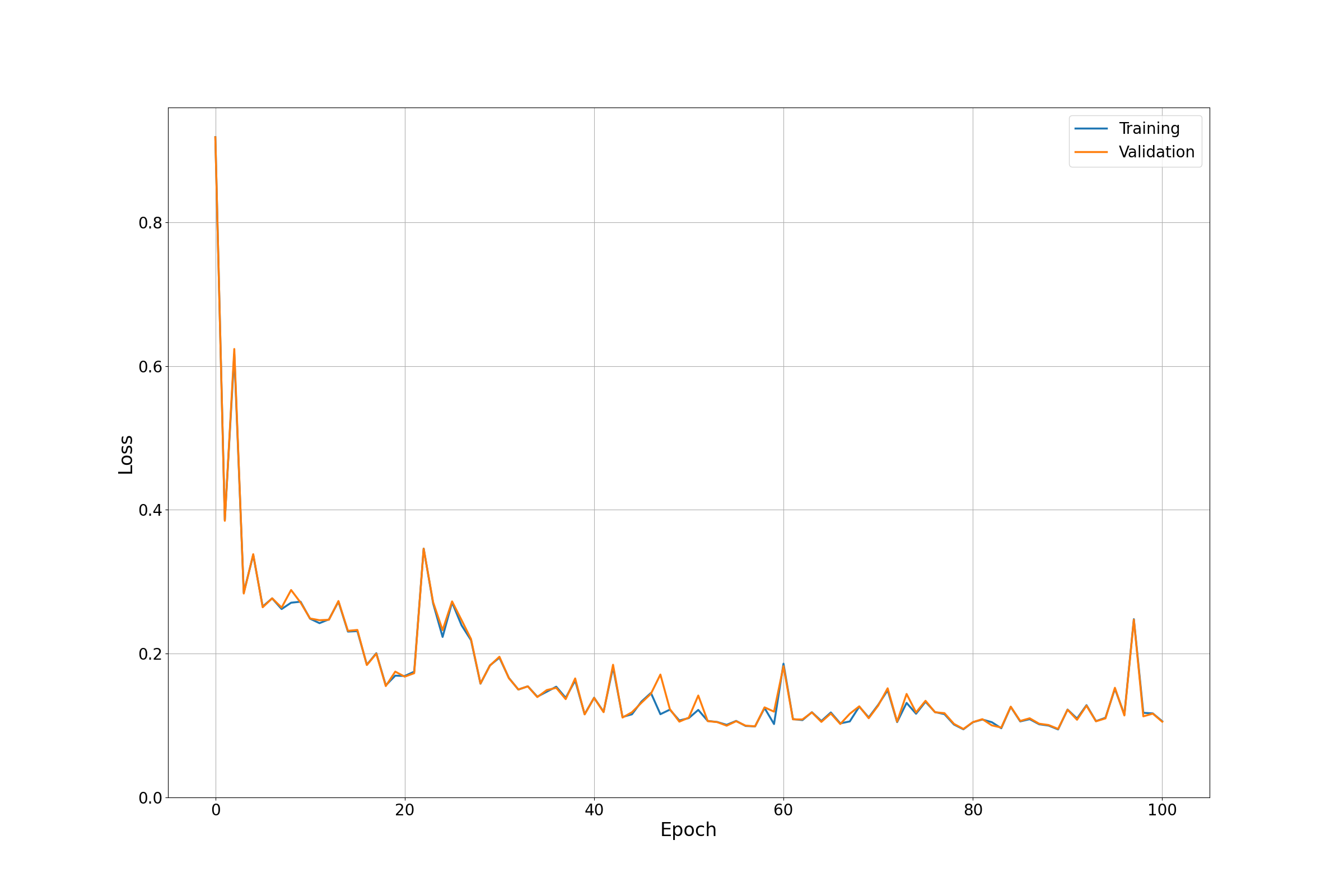}
    \caption{Loss plots of DeepSet training and validation at each epoch for the selected DNN model. The minimal loss is at epoch 77. The selected model yields a testing loss of 0.09629.}
\end{figure}
\begin{figure}[h]
    \centering
    \includegraphics[width=\linewidth,height=0.3\textheight,keepaspectratio=true]{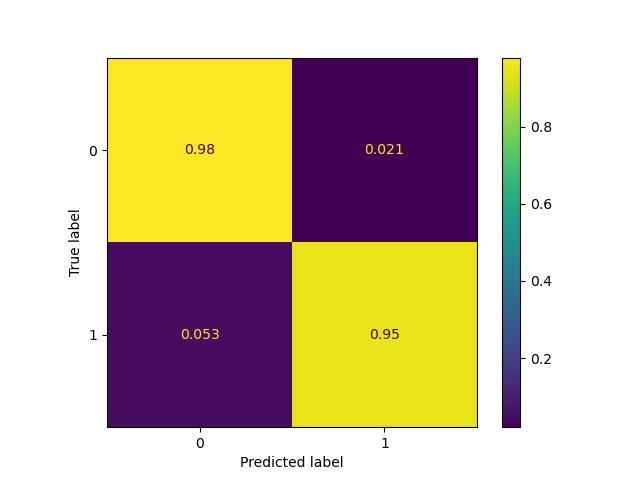}
    \caption{Confusion matrix for the DNN testing with a threshold of 0.5. The resulting accuracy value (defined as the rate of total correct predictions) is 96.34\%. The resulting precision value (defined as the ratio between the number of true signal predictions and the number of total signal predictions) is 97.87\%.}
\end{figure}

\acknowledgments

The work of L.B is supported by an ERC STGgrant (‘BoostDiscovery’, grant No.945878). We thank Prof. David Horn for useful discussions.







\bibliography{main}
\bibliographystyle{JHEP}

\end{document}